# Nucleating Acicular Ferrite with Galaxite ($Al_2O_3MnO$) and Manganese Oxide ($MnO_2$) in FeMnAlC steels

MC McGrath, DC Van Aken, K Song

ABSTRACT

Effectiveness of galaxite ($Al_2O_3MnO$) and manganese oxide ($MnO_2$) in nucleating acicular ferrite was demonstrated in a steel with composition of Fe-13.92Mn-4.53Al-1.28Si-0.11C. The hot rolled and quenched steel had a duplex microstructure of δ-ferrite and partially transformed austenite. Thermal treatments were used to manipulate the concentration and type of oxides and the ferrite plate density was found to correlate with inclusions of low misfit. Both bainitic and acicular ferrite were observed and the prior austenite had a grain diameter of 16.5 μm. Upon tensile testing the retained austenite transformed to α–martensite. Ultimate tensile strength and elongation were 970 MPa and 40%. The yield strength was found to be dependent on the ferrite plate size, which varied linearly with the strength of the austenite. Ductility decreased as the strength disparity between δ-ferrite and deformation-induced α-martensite increased.

## I. INTRODUCTION

Interest to develop new advanced high strength steel for automotive applications is a result of increasing CAFE (corporate average fuel economy) standards where by 2025 an average fuel economy of 54.5 mpg is expected. Automotive steels will need to have ultimate tensile strengths greater than 900 MPa to remain competitive with magnesium and aluminum when these automotive materials are compared on a strength to density (specific strength) basis.[1] Two approaches are being used to increase the specific strength of steel: increasing strength to contain larger proportions of martensite and decreasing density through the addition of aluminum. Dual phase steels with strengths greater than 900 MPa have been achieved by increasing the volume fraction of martensite and austenite (MA). Addition of aluminum to lower the density of steel is also well documented.[2-5] The aluminum atom is 12.6% larger than that of iron and 48% the atomic mass of iron. Thus, steel can be lightweighted by a combination of lattice dilatation and mass reduction. Frommeyer et al.[2] have shown that weight reductions as great as 16-17% can be obtained using aluminum contents of 12 wt.%, but addition of aluminum may also lead to the formation of ordered intermetallic compounds such as FeAl (B2) or $Fe_3Al$ ($DO_3$),[6] which can reduce ductility. Large additions of manganese, greater than 15 wt.%, and carbon up to 1wt.% are often employed to stabilize the austenitic phase. These austenitic high manganese and aluminum steels are referred to as second generation advanced high strength steels and a twin-induced plasticity (TWIP) grade is now commercially produced by POSCO for automotive structures.

Development of third generation advanced high strength steels (AHSS) has been initiated to find lower cost steels than the austenitic second generation AHSS, but with properties exceeding those of first generation AHSS that are dual phase or martensitic.[7] Target properties considered break-through for these new automotive steels would be combinations of ultimate tensile strength and elongation to failure of 1000 MPa and 30% or 1500 MPa and 20%.[8] Matlock et al.[7] using a mechanics approach have shown that a combination of martensite and austenite would be capable of achieving the targeted goals and the quench and partition process appears to be a viable approach.[9] Acicular ferrite and austenite microstructures are also a possibility for meeting the targeted properties;[10-12] however, the nucleation of intragranular ferrite or acicular ferrite is thought to be restricted by grain size and a critical austenite grain size of greater than 20-35 μm has been reported.[13,14] Intergranular nucleation of bainitic ferrite occurs in preference to acicular ferrite at these finer grain structures.[13-18] It should be noted that the structure of these transformation products are identical and the nomenclature relates to where the ferrite is nucleated. Experimental studies have shown that acicular ferrite is bainite and heterogeneously nucleated within the austenite grain i.e. intragranular.[19] The term acicular ferrite is reserved for those plates of ferrite nucleated within the austenite grain.

Acicular ferrite is important for achieving good combinations of strength and ductility.[20,21] Nonmetallic inclusions that are chemically active or have low misfit with ferrite are considered effective nucleation sites for acicular ferrite.[22-26] Some nonmetallic inclusions like manganese sulfides (MnS) are chemically active and deplete manganese from the surrounding austenite, which promotes the nucleation of acicular ferrite.[22,27-32] Park et al.[33] observed galaxite ($Al_2O_3MnO$), aluminum oxide ($Al_2O_3$), aluminum nitride (AlN), manganese sulfides (MnS), and complex aluminum and manganese-rich inclusions formed in an austenitic steel with 3 wt% aluminum and 10 wt% manganese. The misfit between galaxite and ferrite is 1.8% and galaxite was reported as a potent nucleation site for acicular ferrite by Mills et al.[26]

Cerium, calcium, and misch metal additions have also been explored to inoculate steels to form acicular microstructures. Bin et al.[34] showed in situ observations of acicular ferrite nucleation on cerium-rich inclusions with low disregistry between ferrite in a 0.9 wt% manganese steel with a prior austenite grain size of 120 μm. Deoxidation and desulfurization of lightweight FeMnAlC steels was investigated to determine the effectiveness of calcium and misch metal additions to inoculate acicular ferrite.[35] The microstructures and inoculation practices of the FeMnAlC steels are presented in Figure 1. Figure 1(a) highlights the correlation of ferrite plate density with the measured density of inclusions having good lattice registry; total inclusion density did not show a direct correlation. The highest acicular ferrite density was obtained in the calcium treated alloy without any argon stirring (process A). Argon stirring was shown to remove the sulfides with low misfit (52% reduction) and process D samples exhibited 34% lower plate density. All of the steels exhibited an equivalent austenite grain size with a mean linear intercept dimension of approximately 15 μm, which is well below a size where acicular ferrite would be observed. Thus, the correlation between ferrite plate density and inclusion density was unexpected. The alloy produced by process E was 90% transformed and showed larger ferrite plate size compared to the other steels even though the density of ferrite plates remained similar. The argon stirred samples also had a high density of galaxite inclusions.

Producing an inoculated automotive steel can present manufacturing difficulties for the steelmaker. Nonmetallic inclusions usually associated with acicular ferrite are produced during solidification and as a result the distribution within the austenite is less than ideal to produce acicular ferrite. This paper investigated the nature of the nonmetallic inclusions that formed in a duplex FeMnAlC steel of composition Fe-13.92Mn-4.53Al-1.28Si-0.11C and their role in nucleating acicular ferrite. The inclusion types and fraction were manipulated with heat treatments to elucidate their role as nucleating agents. Mechanical testing was performed to demonstrate that a duplex ($\delta$+$\gamma$) steel containing a combination of bainitic and acicular ferrite is capable of achieving the targeted goals for 3$^{rd}$ generation AHSS and at a reduced density.[10,35]

## II. EXPERIMENTAL PROCEDURE

A steel with composition of Fe-13.92Mn-4.53Al-1.28Si-0.11C was produced to study the effectiveness of galaxite and manganese oxide in nucleating acicular ferrite. High purity induction iron, electrolytic manganese, aluminum, ferrosilicon and carbon were melted in a 45 kg (100 lb) induction furnace under an argon protective atmosphere. Deslagging was performed with a low-density granular coagulant. Plates were cast into phenolic no-bake olivine sand molds designed to produce 12.5 cm x 6 cm x 1.7 cm blocks. Chemical analysis was performed by ion coupled plasma spectrometry after sample dissolution in perchloric acid. Nitrogen, phosphorus, and sulfur concentrations were 450 ppm, 170 ppm, and 110 ppm, respectively.

Cast materials were homogenized at 1373 K (1100 ºC) for 2 hours before being air cooled to ambient temperature. The castings were subsequently milled to produce 13.6 mm x 126 mm x 50 mm plates for hot rolling. Hot rolling was performed by reheating the machined castings to 1173 K (900 ºC) and incrementally reducing the ingot by rolling, i.e. starting at 1173 K (900 ºC) with reheating between reductions when the temperature fell below 973 K (700 ºC). The plates were reduced 80% to a thickness of 2.8 mm. After the final rolling pass the plates were either water quenched (method A) or reheated to 1173 K (900 ºC) for 10 minutes prior to water quenching (method B). Specimens from each method were isothermally held at 1123 K (850 ºC) for 2 hours and 47 hours and then water quenched. An additional heat treatment, 15 minutes at 1373 K (1100°C) and water quenching, was performed on a method B specimen that was isothermally held 47 hours at 1123K (850°C). High temperature heat treatments were performed on samples sealed under vacuum in quartz tubes.

Light optical microscopy was used to characterize the microstructures. Standard metallographic practices were employed to polish specimens and each was etched with 2% nital and subsequently etched with 10% sodium metabisulfate to contrast the differences between ferrite and retained austenite in the microstructure. Ferrite plate density was determined based on ASTM E 112-96.[36] Five representative microphotographs at 500x were used from each sample with ten different measurements from each photograph to determine plate density. The α-ferrite plate density included both bainite and acicular ferrite since the fine austenite grain structure made it difficult to distinguish intergranular ferrite (bainite) from intragranular ferrite (acicular). The ferrite plate size was determined as the mean linear intercept measured on random sections. X-ray diffraction was utilized to identify the presence of κ-carbide. Characterization was performed with a Scintag 2000 diffractometer using CuKα radiation. Scans were run from 30-80 degrees with a scan step size of 0.03 degrees. An FEI Tecnai F20 scanning/transmission electron microscope (S/TEM) operated at 200 kV was used to compare the dislocation structures developed from the two hot rolling methods. The S/TEM samples were prepared with an FEI Helios NanoLab 600 FIB/FESEM and specific regions in the samples were cross-sectioned.

Nonmetallic inclusions were characterized by size and composition with automated feature analysis (AFA) using a scanning electron microscope controlled with ASPEX PICA 1020 software. An accelerating voltage of 20 keV and a working distance of 20 mm were used. Samples were metallographically prepared, not etched, and characterized in the short-longitudinal plane of the hot rolled plate. Contrast thresholds, magnification, spot size, electron beam size, scanning speed and image size were kept constant for all samples. For each sample, the AFA application was applied to two different regions of approximately 4 $mm^2$ each. The AFA is limited to inclusions between 0.5 μm and 30 μm in diameter.

Tensile test specimens were machined from the hot rolled products in accordance to ASTM E8-08[37] with a gage length of 50 mm and width of 12.5 mm. Tensile tests were performed with the load axis parallel to the rolling direction. Tests were conducted at room temperature using a displacement rate of 0.01 mm/s. Microhardness measurements were performed with an applied 98.1 mN load for 5 seconds. All of the uncertainties reported in this paper are given at a 95% confidence level.

## III. RESULTS

Hot rolled microstructures of the steel processed by the two methods are shown in Figure 2. Both methods produced microstructures consisting of primary δ-ferrite stringers aligned parallel to the rolling direction, retained austenite, and a combination of bainitic and acicular ferrite plates within the austenite. The density and size of the ferrite plates varied between methods A and B with method A having a greater density of ferrite plates. Figure 3 compares the x-ray diffraction (XRD) patterns obtained from the plates processed by methods A and B. Method B processed plates showed {111} κ-carbide diffraction intensity while no evidence of κ-carbide was observed in the diffraction pattern from the plate processed by method A. Diffraction peak broadening, as exhibited in Figure 3(a) for method A, suggests that recrystallization was suppressed by the low finishing temperature and that the microstructure retained some level of cold work. S/TEM images of representative dislocation structures in δ-ferrite of the hot rolled plates processed by methods A and B are shown in Figure 4. Polishing artifacts from ion milling during specimen preparation were observed as dark features in the δ-ferrite and S/TEM images were used to show greater contrast of the dislocation density. The dislocation density is qualitatively observed to be greater in the plate processed by method A and confirms that some level of cold work was retained. The dislocation structure in the plate processed by method B consisted of small angle boundaries with long straight dislocations distributed in a regular network in the δ-ferrite (Figure 4 (b)).

The hot rolled products were subsequently heat treated at 1123 K (850 ºC) for 2 hours and 47 hours and these microstructures are shown in Figure 5. A quantitative comparison of the α-ferrite plate density, α-ferrite plate size, and volume fractions of austenite and δ-ferrite in the processed materials is shown in Table 1. The α-ferrite density was 53,300 ± 10,900 $mm^{-2}$ in the plate processed by method A. In contrast, the α-ferrite density decreased almost 60% to 21,600 ± 6,000 $mm^{-2}$ when the hot rolled plate was processed by method B and the α-ferrite plate size increased from 0.75 ± 0.19 µm to 1.01 ± 0.32 µm (see Table 1). Heat treating the plates at 1123 K (850 ºC) also produced a 60% decrease in the α-ferrite plate density while the α-ferrite plate size increased with increased holding time at 1123 K (850 °C). There was no statistical difference in the volume fraction of austenite or δ-ferrite between the two plates; however, the austenite measured in the plate processed by method B was inhomogeneous with slightly higher volume fractions of retained austenite in some measured regions. Figure 6 shows that the {111} κ-carbide peak was present in the XRD patterns for all heat treated products.

Nonmetallic inclusions in the various hot rolled and heat treated products were analyzed and the average inclusion diameters are summarized in Table 2. The inclusion size in the method A plate was significantly less compared to the method B plate; the inclusion size in the heat treated plates for both methods were all indistinguishable statistically. Specifically of interest to this study, the average diameters of galaxite and manganese oxides increased from 0.6 µm and 0.7 µm from method A to 1.3 µm and 1.1 µm for method B, respectively. Process method did not appear too affect the average diameter of aluminum nitrides, which measured 4.5 µm.

Figure 7 compares inclusion density and inclusion volume fraction for the hot rolled products and the heat treated plates. The total inclusion density and volume fraction of the hot rolled plate processed by method A were 400 $mm^{-2}$ and 1 x $10^{-3}$, respectively. The inclusion density decreased to 200 $mm^{-2}$ and the volume fraction increased to 1.3 x $10^{-3}$ when the plate was processed by method B. The nonmetallic inclusion density decreased after isothermal heat treatment at 1123 K (850 ºC) as shown in Figure 7. The increase in inclusion volume fraction may suggest that the measurable inclusion content increased as a result of coarsening rather than by additional precipitation. The majority of the inclusions analyzed in the method A plate were galaxite ($Al_2O_3MnO$, 140 $mm^{-2}$) and a combination of aluminum silicates and silica surrounding alumina (Al-Si-O, 85 $mm^{-2}$). On the other hand, the method B plate contained a low density of galaxite (20 $mm^{-2}$) and a high density of aluminum nitrides (70 $mm^{-2}$ vs. 27 $mm^{-2}$). The volume fraction of galaxite decreased to 2 x $10^{-5}$ from 6 x $10^{-4}$ and the volume fraction of nitrides increased 60% to 8 x $10^{-4}$ from 5 x $10^{-4}$ in the method A plate that was isothermally held at 1123 K (850 ºC) for 47 hours. In contrast, the volume fraction of nitrides was not affected by heat treating the plate processed by method B at 1123 K (850 ºC).

Figure 8 (a) displays a light optical micrograph of a plate processed by method B, subsequently isothermally treated at 1123 K (850 ºC) for 47 hours, water quenched, and then heat treated at 1373 K (1100 ºC) for 15 minutes prior to being water quenched. In the 1373 K (1100 ºC) heat treated plate, the α-ferrite plate density increased to 31,300 ± 5,300 $mm^{-2}$ from 7,600 ± 2,200 $mm^{-2}$ measured when the sample was held at 1123 K (850 ºC) for 47 hours. The α-ferrite plate dimension also decreased by 67% to 1.2 μm from 3.65 μm (see Table 3). In addition, the volume fraction of the δ-ferrite decreased to 0.20 ± 0.02 in the 1373 K (1100 ºC) heat treated plate from 0.25 ± 0.04 as observed in the 1123 K (850 ºC) isothermally held plate. Figure 8 shows κ-carbide was present after the 15 minute hold at 1373 K (1100 ºC) and water quenching the sample. Diffraction intensity for {220} κ-carbide in addition to the {111} was observed (see Figure 8 (b)).

Nonmetallic inclusion analyses (see Figure 9) revealed a 46% reduction in the total inclusion volume fraction when the plate was heated to 1373 K (1100 ºC) relative to the method B plate held at 1123 K (850 ºC) for 47 hours. The total volume fraction of nitrides in the 1373 K (1100 ºC) heat treated plate was reduced by 40%. The total volume fraction of the oxides remained relatively constant even though the volume fraction and number density of galaxite increased after heat treatment. The volume fraction of oxides containing silicon decreased to 2 x $10^{-5}$ from 1 x $10^{-4}$.

Microhardness measurements of δ-ferrite and regions of α-ferrite and austenite are compared in Table IV for steels processed by method A, method B, and method A heat treated for 2 hours at 1123 K (850 ºC). Hardness of the austenite region was 9.1% higher than δ-ferrite when the plate was processed by method A. In contrast, the austenite region was 8.3% softer than the δ-ferrite in the plate processed by method B. The microhardnesses of the austenite and δ-ferrite decreased from 431 and 395 to 265 and 326 after a plate processed by method A was heat treated for 2 hours at 1123 K (850 ºC).

Tensile test results for the materials listed in Table IV are shown in Figure 10 and Table V provides a summary of the mechanical properties including strain hardening exponents measured between strain levels of 0.1 and 0.15. The sample processed by method A exhibited discontinuous yielding (649 MPa upper yield stress and 637 lower yield stress) and yield point elongation for up to 12% strain prior to rapid work hardening. Serrated stress flow was also observed during strain hardening of the method A sample. When the product was processed by method B the work hardening behavior appeared linear after yielding at approximately 400 MPa and the strain hardening exponent was 27% lower than the method A (see Table V). A plate processed by method A and heat treated for 2 hours at 1123 K (850 ºC) showed similar tensile behavior to the method B prepared specimen with linear work hardening behavior after yielding (305 MPa) and an absence of serrated flow stress during work hardening. Light optical microscopy revealed α-martensite in the tensile test specimens as shown in Figure 11. Table IV shows the hardness values measured of the martensite was approximately 530 HV.

## IV. DISCUSSION

A combination of bainite and acicular ferrite was observed in the microstructures of the Fe-13.92Mn-4.53Al-1.28Si-0.11C steel studied. Literature shows that acicular ferrite rather than bainite is obtained when inclusion density increases and grain size increases (i.e. the fraction of grain boundary nucleation sites decrease).[13-18] In the work presented here the austenite grain size is limited to the spacing of the δ-ferrite stringers, which measured 16 to 17 μm and remained statistically constant within 1.5 to 2.5 μm after the various isothermal heat treatments as shown in Table VI. Thus, the austenite grain size is eliminated as an experimental variable in the thermal processes investigated.

Thermodynamic predictions for nonmetallic inclusion formation in the steel composition studied are shown in Figure 12. Equilibrium solidification predictions (Figure 12 (a)) show galaxite ($Al_2O_3MnO$) is not expected to form.[38] However, when accounting for solute partitioning during solidification using a method of Scheil segregation modeling, galaxite and silica become favorable in the last 15% of the liquid to solidify (Figure 12 (b)). FactSage[38] was used to predict that the final liquid transformed to austenite and contained lower aluminum concentration. For this last portion of austenite, galaxite is predicted to be stable and should react with nitrogen and silicon in solid solution to form $SiO_2$ and AlN at temperatures approximately below 1133 K (860 ºC). Figure 13 (a) shows verification of $SiO_2$ nucleating on an aluminum oxide particle in a method A plate heat treated at 1123 K (850 ºC). These thermodynamic calculations agree with the observed reduction in galaxite particles and an increase in silicon-aluminum oxides during heat treatment at 1123 K (850 ºC). Galaxite was nearly absent ($f = 9.2 \times 10^{-6}$) in the method B processed plate after 47 hours at 1123 K (850 ºC). In contrast when the 47 hour heat treated plate was reheated to 1373 K (1100 ºC) the volume fraction of galaxite increased to $2.5 \times 10^{-5}$. Galaxite was cited as being an effective nucleation site for acicular ferrite[26] and many of the austenite grains showed evidence of ferrite nucleation towards the center of the grain as shown in Figure 13 (b). However, it is difficult to claim intragranular nucleation when the austenite grain size is less than 20μm.

Different mechanisms have been proposed to explain the role of nonmetallic inclusions in nucleating acicular ferrite.[22-32, 39,40] Recent literature[34, 41] appears to reach a consensus that inclusions are chemically active and cause manganese solute depletion, which reduces the stability of the austenite; and these inclusions generally have a small lattice disregistry with the ferrite which would reduce the interfacial energy. Literature on acicular ferrite nucleation suggest that a nonmetallic inclusion with a misfit less than 6% is an effective nucleation site for ferrite; an inclusion with a misfit between 6-12% is moderately effective; and an inclusion with greater than 12% misfit would not be expected to contribute to nucleation.[25] Table 7 lists the misfit of inclusions that are probable nucleation sites for acicular ferrite.[23, 26] Barbaro et al.[14] noted that curvature of larger inclusions increased the probability to nucleate acicular ferrite due to the decrease in the interface curvature between the ferrite and inclusion interface to provide better epitaxy. The potency of the inclusion as a nucleation site is not linearly dependent on the inclusion size. An inclusion with a diameter greater than 0.6 μm had a higher probability to nucleate acicular ferrite even though an inclusion with a diameter of 0.4 μm was noted as a possible nucleation site for acicular ferrite. Inclusions less than 0.6 μm in diameter were omitted from the analyses discussed below.

Figure 14 tests the relationship between the density of α-ferrite and the density of nonmetallic inclusions listed in Table VII. Densities of manganese oxide (including manganese oxides nucleated on manganese sulfides, Figure 14 (a)) and galaxite (Figure 14 (b)) suggest a direct correlation with the density of ferrite plates. These inclusions have less than 2% lattice misfit with α-ferrite (see Table VII). Galaxite and manganese oxide were cited in past studies as having a low misfit with ferrite, which led to acicular ferrite nucleation.[23, 26] Manganese oxide was also cited as depleting solute from the austenite and would thus promote the nucleation of acicular ferrite.[42,43] In this study galaxite and manganese oxide (including these same oxides nucleated on manganese sulfides) were shown to have a strong correlation with the α-ferrite density and therefore are viable nucleation sites for ferrite.

In contrast there was no relationship between the α-ferrite density and the densities of aluminum oxide and manganese sulfide as shown in Figures 14 (c) and 14 (d). In literature manganese sulfide was reported as a possible nucleation site for acicular ferrite. A manganese depleted region surrounding an inclusion was hypothesized to catalyze the ferrite nucleation.[27-32] Mabuchi et al.[27] showed manganese depleted zones surrounded manganese sulfides but these regions were homogenized after the completion of precipitation in a steel containing 1.4 wt% manganese. The manganese sulfides were ineffective nucleation sites for acicular ferrite after the elimination of the manganese depletion zone.[27] In the study presented here, the volume fraction and density of manganese sulfides remained constant after the heat treatments at 1123 K (850 ºC) for 2 and 47 hours and we suggest that any manganese depleted zone would be eliminated by the 47 hour heat treatment time. However, no difference in correlation between heating time and ferrite plate density was observed and all measurements fell on the same trend line. The manganese profile across the interface of manganese sulfide and austenite in a sample heat treated at 1373 K (1100 ºC) was acquired using energy dispersive spectroscopy in a S/TEM. An TEM image in Figure 15 (a) shows morphology of the manganese sulfide and the austenite grain was contrasted dark since the sample was tilted to satisfy a <100> zone axis. Figure 15 (b) shows austenite was not depleted of manganese near the interface, which would support the observation that manganese sulfides were ineffective nucleation sites for acicular ferrite nucleation. The same should be true for the galaxite and manganese oxides, but the correlation between ferrite plate density and oxide number density remains strong even after 47 hours of isothermal heat treatments suggesting that lattice registry is playing a more important role.

Figure 16 shows a strong correlation between the ferrite plate density and the density of inclusions determined viable based on Figure 14. The relationship determined was $\rho^{\alpha} = 440\rho^{inclusion} + 2{,}800$, where $\rho^{\alpha}$ is the α-ferrite density and $\rho^{inclusion}$ is the viable inclusion density. The relationship developed from Figure 16 suggests that the bainitic ferrite contribution was 2,800 $mm^{-2}$. The austenite grain size was approximately 16.5 μm so there would be approximately 2,810 grains/$mm^2$ and this corresponds to an average of 1 bainite plate per grain.[44] Additionally this relationship suggests that a viable inclusion greater than 0.6 μm will cause a nucleation event for 440 acicular ferrite plates/$mm^2$ per inclusion. However, the measured inclusion density is much lower than the austenite grain density and correlates to less than 1 inclusion (greater than 0.6 μm) per grain. Inclusion diameters less than 0.5 μm are difficult to detect with the instrument used for automated feature analysis and sectioning effects may have led to a smaller than actual measured inclusion density. Inclusions less than 0.6 μm may also have contributed to the nucleation of acicular ferrite, but to a lesser effect based upon the work of Barbaro et al.[14]

The density of acicular ferrite plates was a magnitude larger than the density of viable inclusions, which is similar to observations made by Barbaro et al.[14] and Ricks et al.[45] suggested nucleation occurred sympathetically on the broad ferrite plate after the inclusion surface became saturated. A recent study by Wan et al.[41] investigated the mechanism for the interlocked microstructure of acicular ferrite using three-dimensional reconstruction techniques. Multiple ferrite plates nucleating on a single inclusion, sympathetic nucleation along the broad face of a previous formed ferrite plate, and hard impingement with intersection between ferrite plates were noted as being responsible for the interlocking ferrite microstructure. Sympathetic nucleation would result in the ferrite density being larger than the inclusion density, which was also observed in the results reported here. The strong correlation between ferrite plate density and inclusions of galaxite and manganese oxide support a conclusion that acicular ferrite can be formed in small grained austenitic structures. Furthermore, this study shows that a steel chemistry can be formulated where the inoculating inclusions can be produced by heat treatment is the solid state rather than relying on their formation during solidification.

Method A produced an ultimate tensile strength of 970 MPa and an elongation to failure of 40% where the tensile strength is slightly below that considered a break-through for a third generation advanced high strength steel. The mechanical properties were significantly lower in the steel processed by method B and in the steel produced by method A after heating for 2 hours at 1123 K (850 ºC). For example, the yield strength decreased to 400 MPa when the plate was processed by method B (method A – 642 MPa) and to 305 MPa when the method A plate was held at 1123 K (850 ºC) for 2 hours.

For steels with acicular microstructures, the yield strength is related to the inverse of the ferrite plate size.[46] Dallum et al.[47] observed coarser acicular ferrite in an HSLA steel when the prior austenite grain size was reduced which resulted in fewer nucleation sites. It was reasoned that the acicular ferrite was coarser due to a reduced nucleation rate. Singh et al.[48] modeled the thickness of ferrite plates in steels with an average composition of Fe-2Si-2Mn-0.25C by accounting for the transformation temperature, composition, and amount of supercooling relative to the transformation temperature. It was shown that the strength of austenite at the transformation temperature and the chemical free energy change were the major contributors to the thickness of ferrite plates. The average size of the ferrite in this study generally increased as the ferrite density decreased (Figure 16 (a)), which was related to fewer viable nonmetallic inclusions. Figure 16 (a) suggests the ferrite size was controlled by a factor other than just the ferrite nucleation rate since two separate trend lines can be drawn for the data. The ferrite plate size data can be reduced to a single trend line by plotting the data as a function of austenite hardness as shown in Figure 16 (b). Here room temperature hardness is used as a proxy for the strength of the austenite at the transformation temperature. Figure 16 (c) shows an inverse linear relationship between the hardness of austenite and the fraction of nitrides. It should be noted that the method A processed material was excluded from these correlations since the material was in a work hardened state prior to transformation. Increased nitrogen in solution, as evidenced by a decrease in volume fraction of nitrides in method A processed steel, would result in solid solution strengthening of the austenite. The increase in the ferrite plate size (method A: 0.75 μm vs. method B: 1.01 μm vs. method A heat treated 2 hours: 3.12 μm) observed in this study may have led to some of the decrease in yield strength. Tensile tests were repeated for a number of the heat treated samples and the yield strength as a function of the ferrite plate width is shown in Figure 17.

The tensile ductility as measured by elongation to failure of the plates processed by method B and method A plates that were heat treated for 2 hours at 1123 K (850 ºC) decreased to 34% and 32%, respectively. The loss in ductility may be explained by considering the change in strength of the various microstructures, δ-ferrite in particular, with processing relative to the hardness of the martensite formed upon deformation. Sun et al.[49] modeled the ductile failure mechanism in dual phase steels and showed that the strength disparity between mechanically stable ferrite phase and deformation-induced martensite phase adversely affects the ductility. A large discrepancy between the ferrite and martensite would lead to early void nucleation and reduced tensile ductility.[49] Hasegawa et al.[50] has also observed minimizing the hardness differences between ferrite and martensite would lead to higher formability in a DP980 steel. The retention of the cold worked dislocation structure in the δ-ferrite when the plate was processed by method A (Figure 4 (a)) and the possibility of strain aging as discussed below led to the higher hardness of the δ-ferrite, decreased the hardness disparity with the martensite, and thus increased elongation.

Serrations in the stress-strain curve and high strain hardening exponents are characteristic of dynamic strain aging[51,52] and were observed only in the steel processed by method A. Serrated flow disappeared and strain hardening exponents decreased from 0.48 to 0.37 when a method A plate was heat treated at 1123 K (850 ºC) for 2 hours. The steel processed by method B also had a reduced strain hardening exponent (0.36) and did not show serrated flow. Mn-C dipoles have been shown to aid in dynamic strain aging and rapid work hardening in high manganese steels like Hadfield steels.[51,53] κ-carbide, $(Fe,Mn)_3AlC$, has been reported to form in duplex ferritic and austenitic steels with aluminum and carbon concentrations ranging from 5 to 7 wt% and 0.1 to 0.4 wt%, respectively.[54] κ-carbide formation was suppressed in the plate processed by method A where the cold worked dislocation structure was remnant. κ-carbide was observed in the heat treated plates. A consequence of κ-carbide formation could be the removal of carbon from the austenite, which would reduce the Mn-C dipole concentration and lead to lower strain hardening exponents. In a study by Zudiema et al.[55] the addition of aluminum to a Hadfield steel caused the dynamic strain aging behavior to be suppressed at room temperature. It has been shown by Medvedeva et al.[56] that Fe-Al-C defect structures with the same coordination as κ-carbide are energetically preferred. This may suggest that deformation processing below the recrystallization temperature may destroy these Fe-Al-C clusters and inhibit the nucleation of κ-carbide as evidence in Figure 3a.

Discontinuous yielding was observed only in the plate processed by method A, which indicates a strain aging behavior. According to literature discontinuous yielding is attributed to interstitial nitrogen in solution.[57-59] A plate processed by method A showed a low volume fraction ($6.2 \times 10^{-4}$) of nitrides which suggest that nitrogen remained in solution. The δ-ferrite had a high hardness and dislocation density, which is prerequisite for nitrogen pinning and strain aging. The volume fraction of nitrides increased to $8.8 \times 10^{-4}$ after a plate processed by method A was heat treated at 1123 K (850 ºC) for 2 hours. The removal of nitrogen from solution and recovery of the dislocation structure resulted in the disappearance of the discontinuous yielding.

## V. CONCLUSION

This study shows that an acicular ferritic steel can be formulated where nonmetallic inclusions can be precipitated by thermal processing rather than formation during solidification. Thermal treatments were used to manipulate the concentration and type of oxides and the ferrite plate density was found to correlate with inclusions of low misfit. Specifically, manganese oxide and galaxite were effective nucleation sites for acicular ferrite in a duplex FeMnAlC steel of composition Fe-13.92Mn-4.53Al-1.28Si-0.11C. The ferrite plate density did not depend on the density of manganese sulfide. The heat treatments employed eliminated any manganese depleted zones next to these sulfides and as a result the MnS was ineffective in nucleating acicular ferrite. However, the same rationale should be true for galaxite and manganese oxides, but the correlation between ferrite plate density and oxide number density remained strong even after 47 hours at 1123 K (850 °C) suggesting that the low lattice disregistry between ferrite and either galaxite or manganese oxide was sufficient to encourage acicular ferrite formation.

Both bainitic and acicular ferrite were observed in austenite with an average grain diameter of 16.5 μm, which was below the previously reported critical austenite diameter for acicular ferrite formation. The strong correlation between ferrite plate density and inclusions of galaxite and manganese oxide support a conclusion that acicular ferrite can be formed in small grained austenitic structures that might be produced for automotive applications.

The duplex steel achieved an ultimate tensile strength and elongation of 970 MPa and 40% when the steel was finished at a low rolling temperature and possessed some degree deformation. Upon tensile testing the retained austenite transformed to α-martensite. Reheating the steel after hot rolling reduced the tensile ductility as a result of a greater strength disparity between the deformation-induced α-martensite and the heat treated microstructure. The yield strength was found to be dependent on the ferrite plate size, which varied linearly with the strength of the austenite.

The strain hardening exponent increased to 0.48 when κ-carbide was suppressed. Rolling deformation below the recrystallization temperature might have destroyed the energetically favorable Fe-Al-C clusters that would have led to nucleation of κ-carbide upon quenching. Discontinuous yielding was also observed when nitride volume fraction was low. Nitrogen in solution along with a high dislocation density would promote the discontinuous yielding of the ferritic structures. Conversely the discontinuous yielding disappeared when the volume fraction of nitrides increased.

## ACKNOWLEDGEMENTS

This work was supported in part by the National Science Foundation (NSF) and the Department of Energy under contract CMMI 0726888. The FEI Helios NanoLab dual beam FIB was obtained with a Major Research Instrumentation grant from NSF under contract DMR-0723128 and the FEI Tecnai F20 scanning/transmission electron microscope was obtained with a Major Research Instrumentation grant from NSF under contract DMR-0922851. The authors gratefully acknowledge the support of the Graduate Center for Materials Research and in particular Eric Bohannen for help with x-ray diffraction. Meghan McGrath was supported by a Department of Education GAANN fellowship under contract P200A0900048.

LIST OF FIGURE CAPTIONS

Fig. 1 – (a) Ferrite plate density decreases with decreasing amount of viable nonmetallic inclusions for nucleation in six different FeMnAlC steels (labeled A through E) with varying deoxidation and desulfurization practices.[15] Deoxidation/desulfurization practices: A- Ca additions; B- no Ce additions; C- Ce additions D- Ca additions and 30s Ar-stirring; and E- calcium additions and 60s Ar-stirring. Microstructures from the steels processed by (b) A, (c) B, and (d) E are shown.[35]

Fig. 2 – Light optical micrographs of Fe-13.92Mn-4.53Al-1.28Si-0.11C plates hot rolled at 1173 K (900 ºC) and were either (a) water quenched immediately (method A) or (b) reheated at 1173 K (900 ºC) for 10 minutes before being water quenched (method B).

Fig. 3 – XRD patterns of hot rolled products (a) water quenched immediately (method A) and (b) reheated for 10 minutes at 1173 K (900 ºC) prior to water quenching (method B).

Fig. 4 – S/TEM images of the dislocation structure of δ-ferrite in the hot rolled plates processed by (a) method A and (b) method B. Polishing artifacts from ion milling during preparation of S/TEM specimens were observed in the δ-ferrite.

Fig. 5 – Light optical micrographs from hot rolled plates that were water quenched immediately and then subsequently heat treated at 1123 K (850 ºC) for (a) 2 hours and (b) 47 hours; and micrographs from plates that were reheated prior to water quenched and then heat treated at 1123 K (850 ºC) for (c) 2 hours and (d) 47 hours. Specimens were water quenched after isothermal holds.

Fig. 6 – XRD patterns of hot rolled products processed by (a) method A then heat treated at 1123 K (850 ºC) for 2 hours; (b) method A then heat treated at 1123 K (850 ºC) for 47 hours; (c) method B then heat treated at 1123 K (850 ºC) for 2 hours; and (d) method B then heat treated at 1123 K (850 ºC) for 47 hours. Specimens were water quenched after annealing. κ-carbide peaks were observed in the patterns of the heat treated plates.

Fig. 7 – (a) Comparison of inclusion density between hot rolled products that were rolled at 1173 K (900 ºC) and processed by method A. Plates processed by method A were then heated at 1123 K (850 ºC) for 2 hours and 47 hours. (b) Comparison of volume fraction of inclusions for the various hot rolled products processed by method A. (c) Comparison of inclusion density between hot rolled products that were processed by method B and then subsequently heat treated at 1123 K (850 ºC) for 2 hours and 47 hours. (d) Comparison of the volume fraction of inclusions in plates processed by method B and subsequently heat treated. Inclusions identified in legend reading left to right and are plotted in the same order from top to bottom.

Fig. 8 – (a) Light optical micrograph of method B plate that was heat treated at 1373 K (1100 ºC) (15 minutes) after heat treating at 1123 K (850 ºC) for 47 hours. (b) XRD pattern showing diffraction intensity for κ-carbide. Sample was water quenched after each heat treatment.

Fig. 9 – (a) Comparison of inclusion density between plates that were hot rolled at 1173 K (900 ºC) and then processed by method B; heat treated at 1123 K (850 ºC) for 47 hours; and heat treated at 1373 K (1100 ºC) for 15 minutes. (b) Comparison of volume fraction of inclusions for the various hot rolled products. Inclusions identified in legend reading left to right and are plotted in the same order from top to bottom.

Fig. 10 – Representative tensile tests of Fe-13.92Mn-4.53Al-1.28Si-0.11C after being processed by method A (black curve); method B (red curve); and method A heat treated for 2 hours at 1123 K (850 ºC) (blue curve).

Fig. 11 – Light optical micrograph from gage section of a plate that was processed by method A and pulled in tension to failure at 970 MPa with 44% elongation.

Fig. 12 – FactSage[38] predictions for nonmetallic inclusions in steels with composition (a) Fe-13.92Mn-4.53Al-1.28Si-0.045N-0.11C-0.05O-0.05S (bulk); (b) Fe-20.3Mn-0.7Al-3.2Si-1.2N-0.25C-0.05O-0.05S (last 15% to solidify).

Fig. 13 – Secondary electron images of (a) silica nucleating on an aluminum oxide particle in a sample processed by method A and heat treated at 1123 K (850 ºC) for 47 hours and (b) galaxite observed within a cluster of acicular ferrite in a sample processed by method B and held at 1373 K (1100 ºC) after being heat treated at 1123 K (850 ºC) for 47 hours.

Fig. 14 – Relationships between acicular ferrite density and density of (a) $MnO_2$ and $MnO_2$ nucleated on MnS; (b) galaxite; (c) $Al_2O_3$; and (d) MnS for inclusions larger than 0.6 μm.

Fig. 15 – (a) TEM image showing boundary between manganese sulfide (MnS) and austenite (FCC). (b) Manganese profile across the MnS and FCC boundary. There was no depletion of manganese.

Fig. 16 – A positive relationship between acicular ferrite plate density and inclusions with low misfit that were larger than 0.6 μm was observed.

Fig. 17 – α-ferrite plate size was a function of (a) ferrite plate density and (b) austenite hardness. (c) The hardness of the austenite was a function of the total nitrides. The datum of processing by method A was removed due to this steel being in a work hardened state prior to transformation.

Fig. 18 – Yield strength was inversely related to the α-ferrite plate size. The plate processed by method A had retention of the dislocation structure from cold working which increased the strength of the material. This data point was excluded from determining the linear relationship between yield strength and the inverse of the size of the acicular ferrite.

LIST OF FIGURES

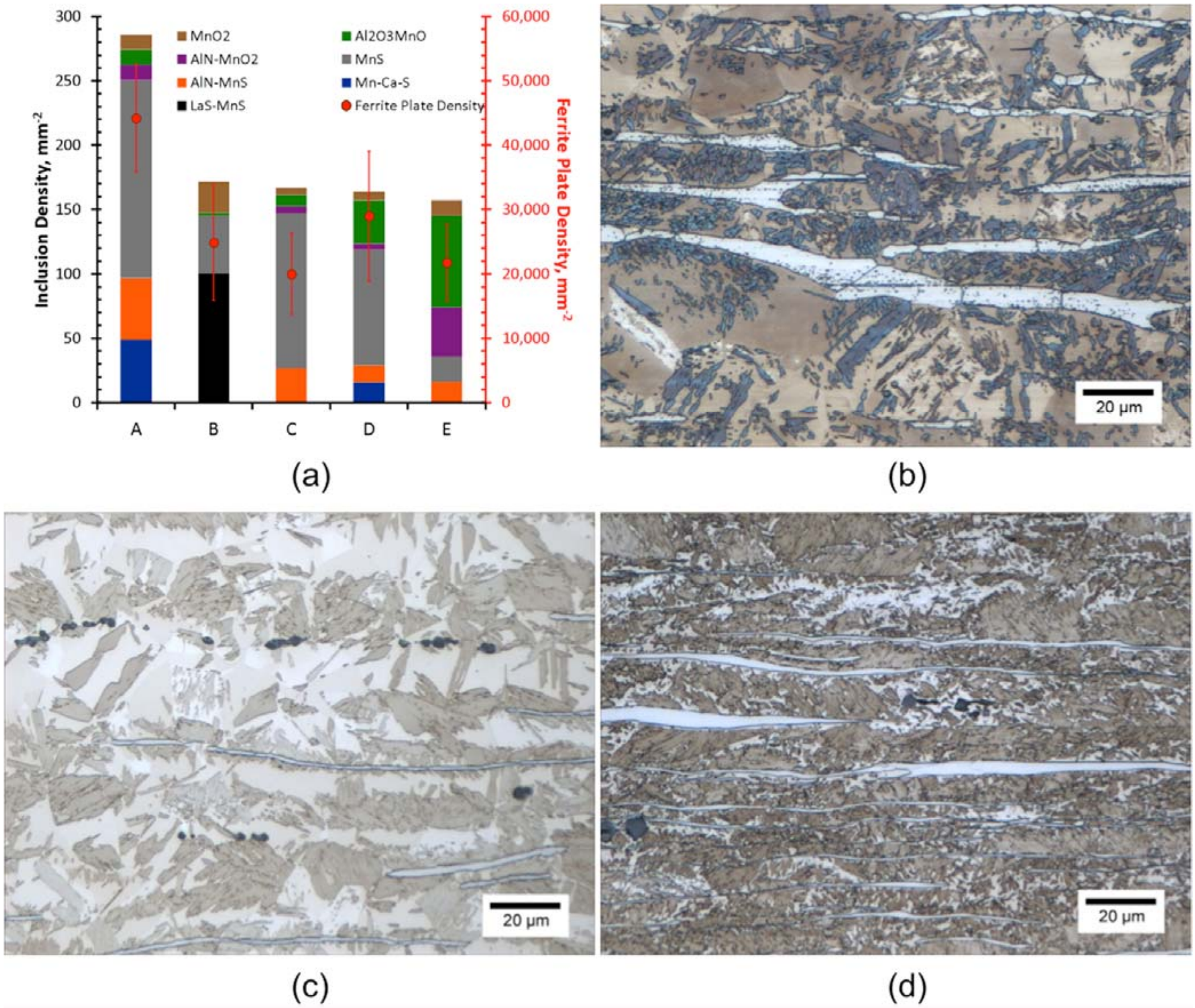

Fig. 1 – (a) Ferrite plate density decreases with decreasing amount of viable nonmetallic inclusions for nucleation in six different FeMnAlC steels (labeled A through E) with varying deoxidation and desulfurization practices.[15] Deoxidation/desulfurization practices: A- Ca additions; B- no Ce additions; C- Ce additions D- Ca additions and 30s Ar-stirring; and E- calcium additions and 60s Ar-stirring. Microstructures from the steels processed by (b) A, (c) B, and (d) E are shown.[35]

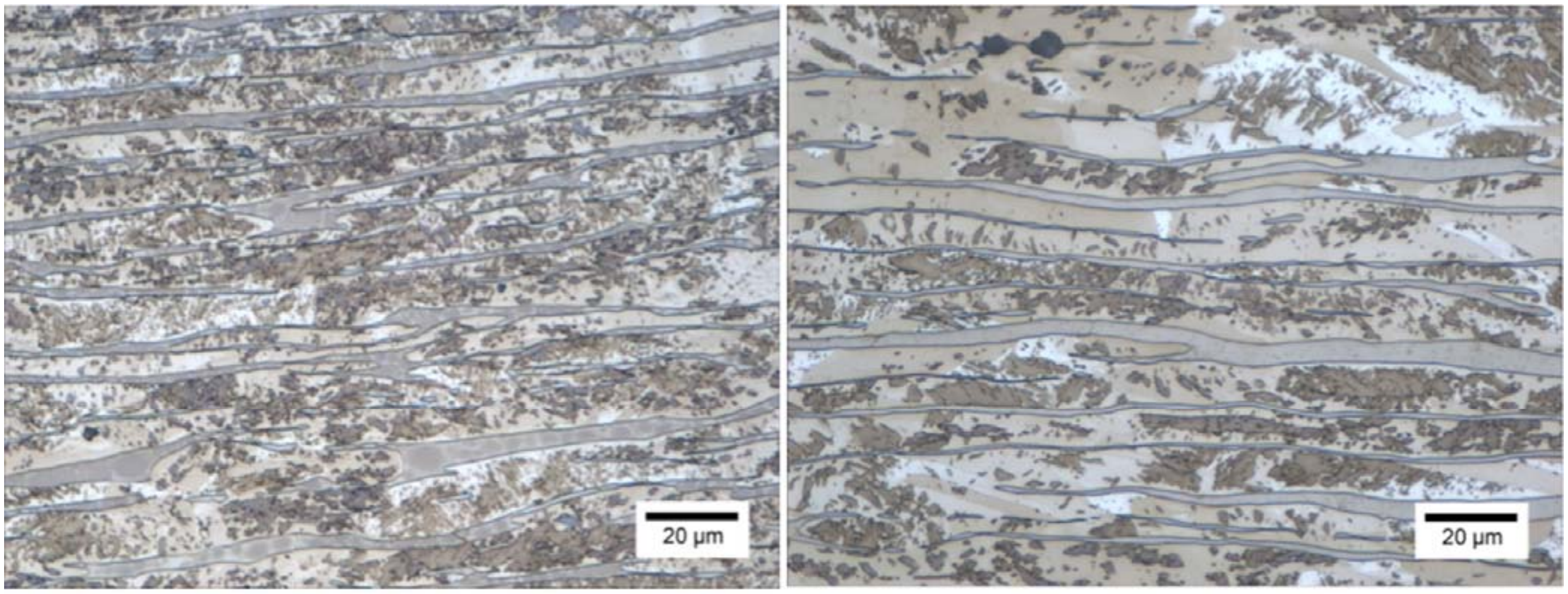

(a) (b)

Fig. 2 – Light optical micrographs of Fe-13.92Mn-4.53Al-1.28Si-0.11C plates hot rolled at 1173 K (900°C) and were either (a) water quenched immediately (method A) or (b) reheated at 1173 K (900°C) for 10 minutes before being water quenched (method B).

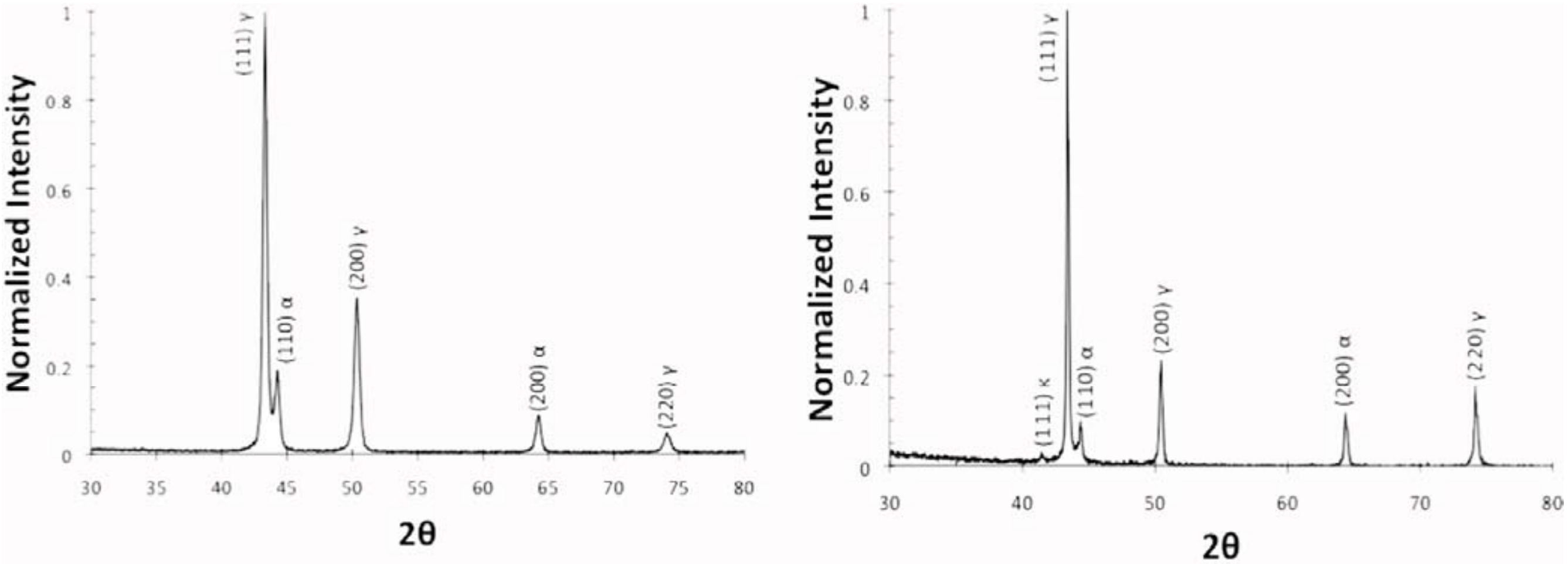

(a) (b)

Fig. 3 – XRD patterns of hot rolled products (a) water quenched immediately (method A) and (b) reheated for 10 minutes at 1173 K (900°C) prior to water quenching (method B).

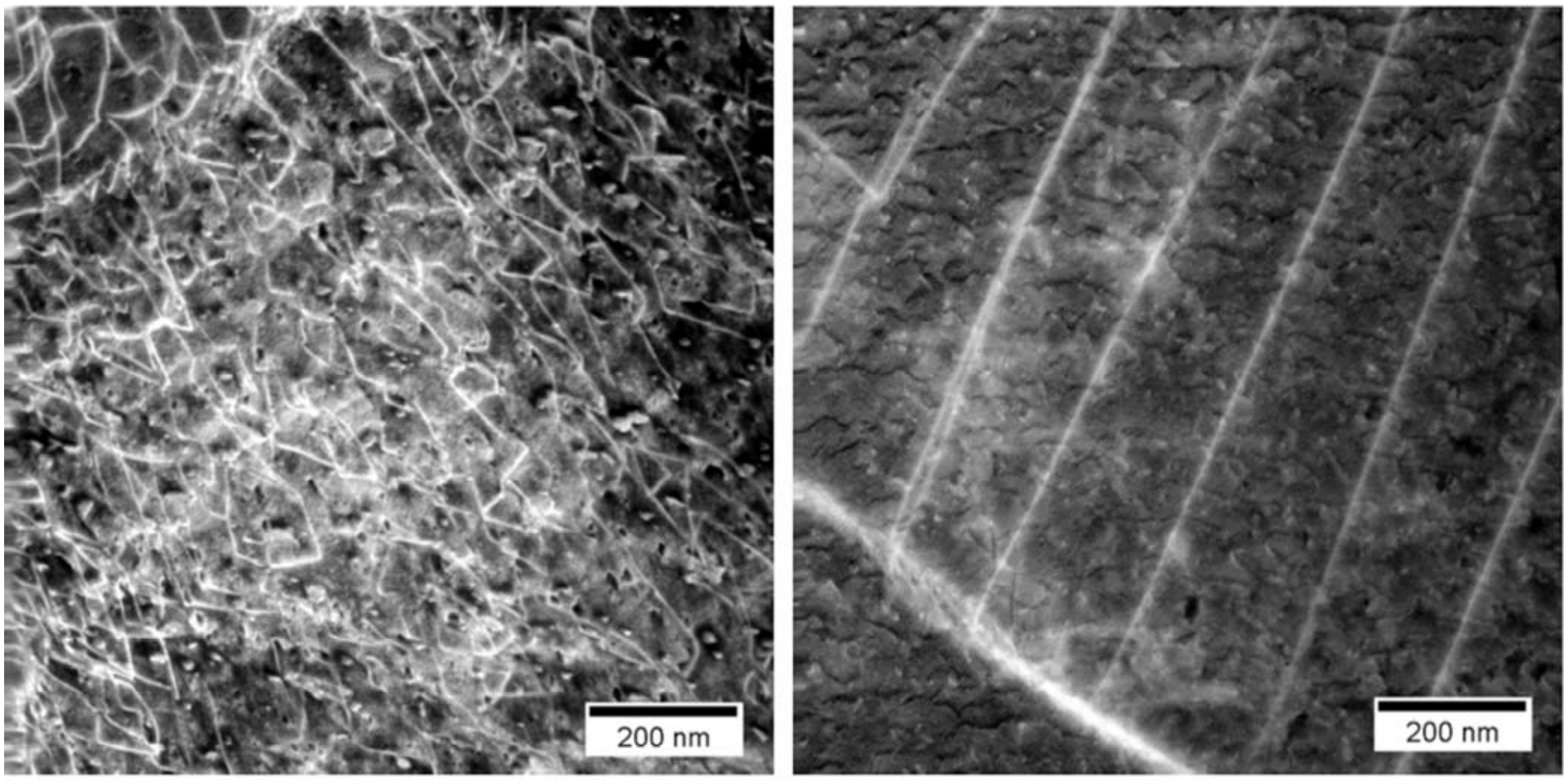

(a) (b)

Fig. 4 – S/TEM images of the dislocation structure of δ-ferrite in the hot rolled plates processed by (a) method A and (b) method B. Polishing artifacts from ion milling during preparation of S/TEM specimens were observed in the δ-ferrite.

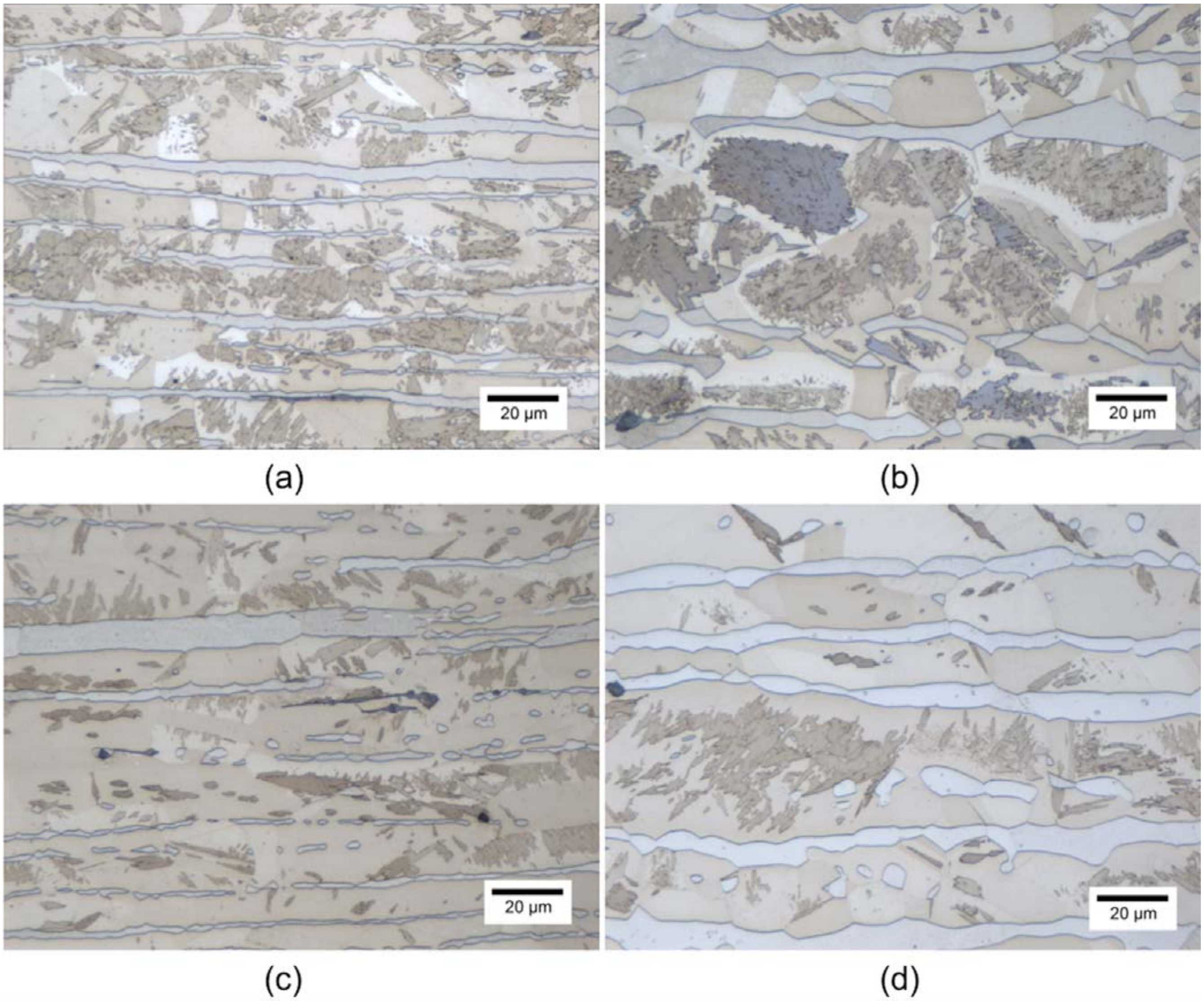

Fig. 5 – Light optical micrographs from hot rolled plates that were water quenched immediately and then subsequently heat treated at 1123 K (850°C) for (a) 2 hours and (b) 47 hours; and micrographs from plates that were reheated prior to water quenched and then heat treated at 1123 K (850°C) for (c) 2 hours and (d) 47 hours. Specimens were water quenched after isothermal holds.

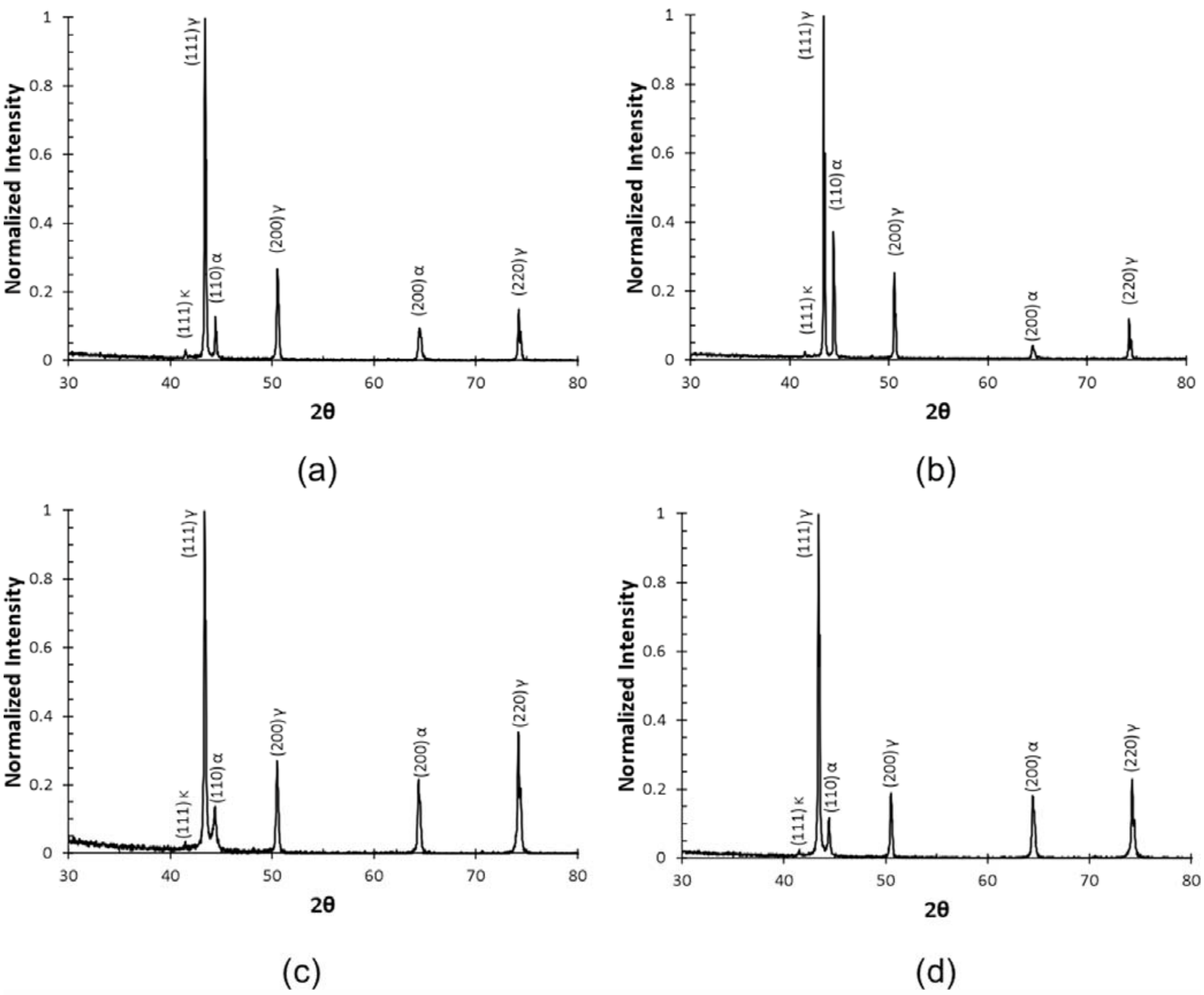

Fig. 6 – XRD patterns of hot rolled products processed by (a) method A then heat treated at 1123 K (850°C) for 2 hours; (b) method A then heat treated at 1123 K (850°C) for 47 hours; (c) method B then heat treated at 1123 K (850°C) for 2 hours; and (d) method B then heat treated at 1123 K (850°C) for 47 hours. Specimens were water quenched after isothermal heat treatments. κ-carbide peaks were observed in the patterns of the heat treated plates.

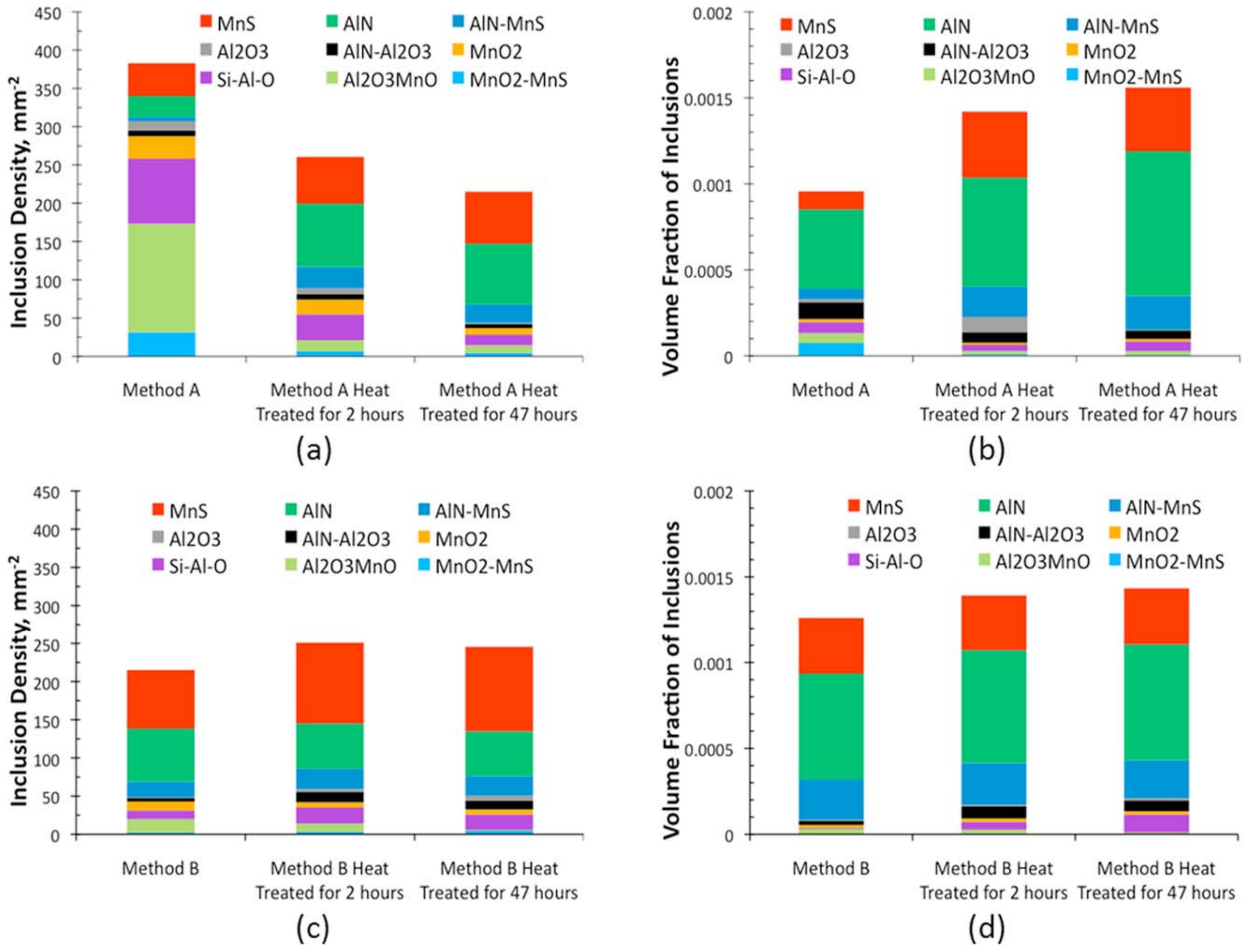

Fig. 7 – (a) Comparison of inclusion density between hot rolled products that were rolled at 1173 K (900°C) and processed by method A. Plates processed by method A were then heat treated at 1123 K (850°C) for 2 hours and 47 hours. (b) Comparison of volume fraction of inclusions for the various hot rolled products processed by method A. (c) Comparison of inclusion density between hot rolled products that were processed by method B and then subsequently heat treated at 1123 K (850°C) for 2 hours and 47 hours. (d) Comparison of the volume fraction of inclusions in plates processed by method B and subsequently annealed. Inclusions identified in legend reading left to right and are plotted in the same order from top to bottom.

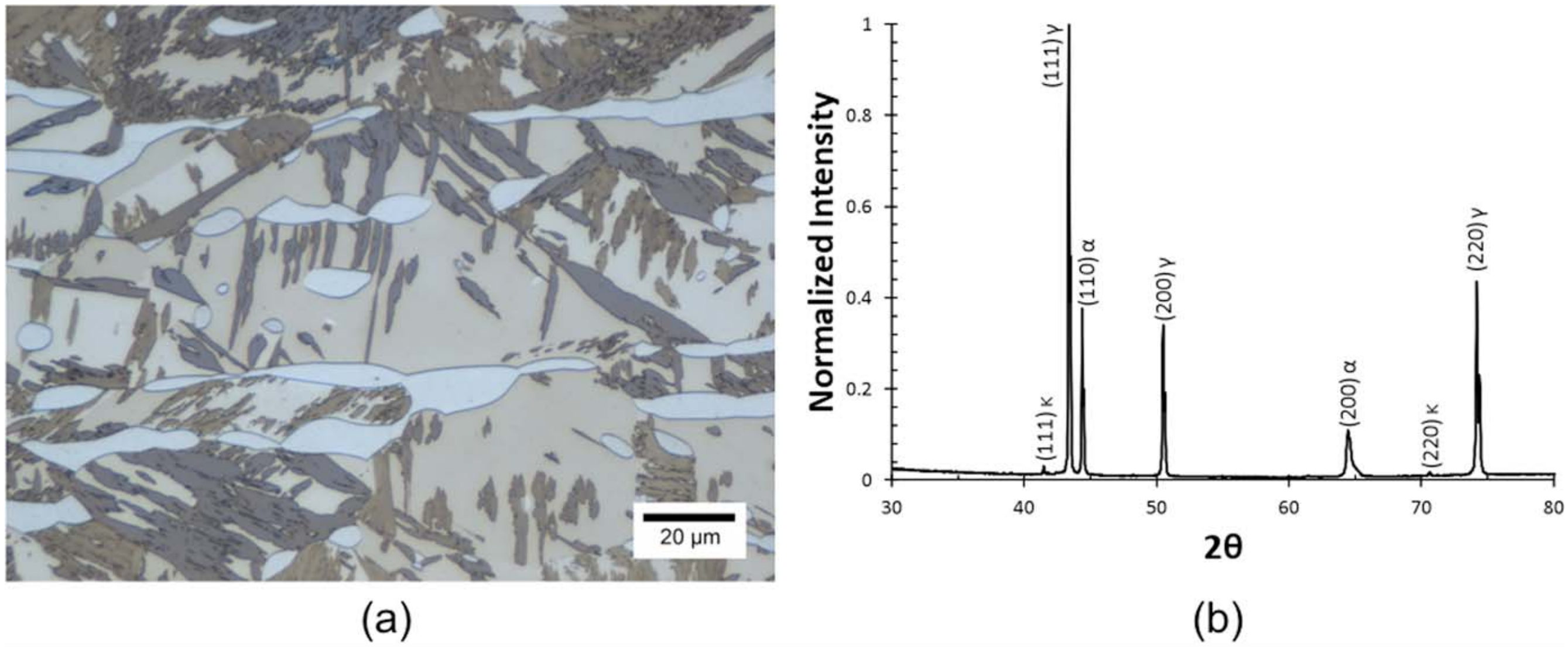

Fig. 8 – (a) Light optical micrograph of method B plate that was heat treated at 1373 K (1100°C) (15 minutes) after heat treating at 1123 K (850°C) for 47 hours. (b) XRD pattern showing diffraction intensity for κ-carbide. Sample was water quenched after each heat treatment.

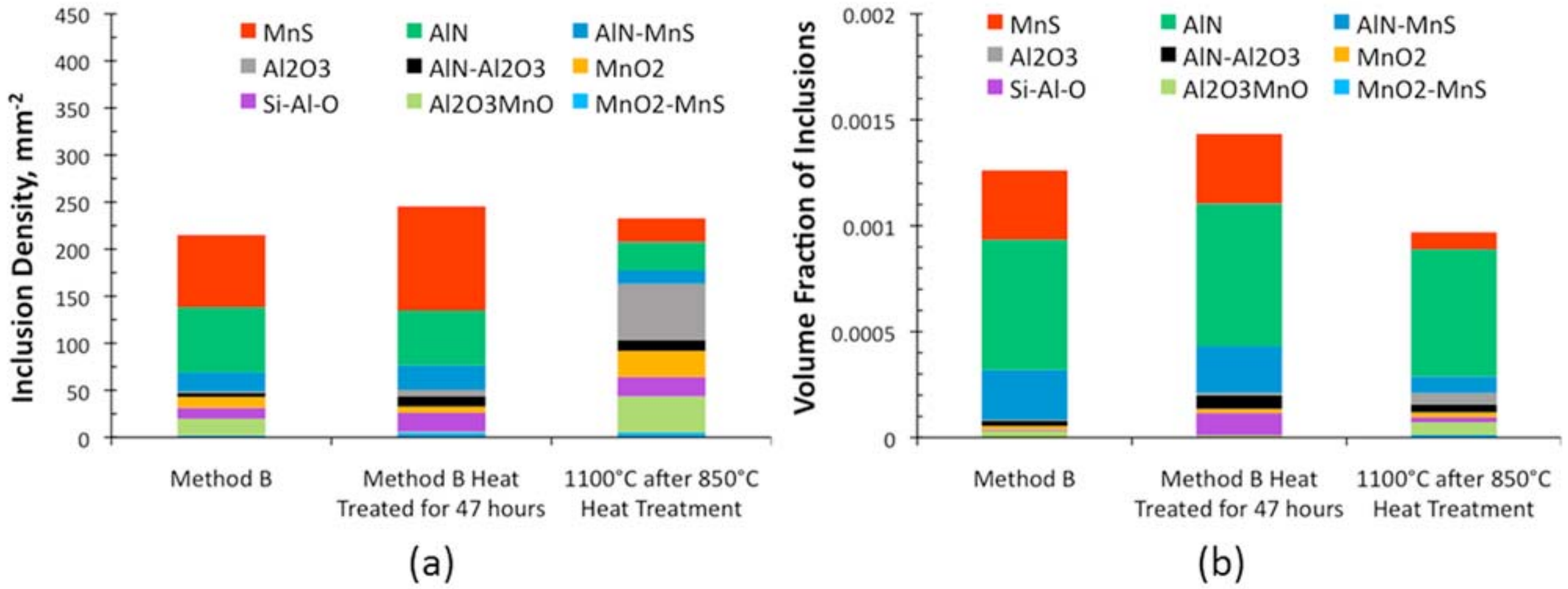

Fig. 9 – (a) Comparison of inclusion density between plates that were hot rolled at 1173 K (900°C) and then processed by method B; heat treated at 1123 K (850°C) for 47 hours; and heat treated at 1373 K (1100°C) for 15 minutes. (b) Comparison of volume fraction of inclusions for the various hot rolled products. Inclusions identified in legend reading left to right and are plotted in the same order from top to bottom.

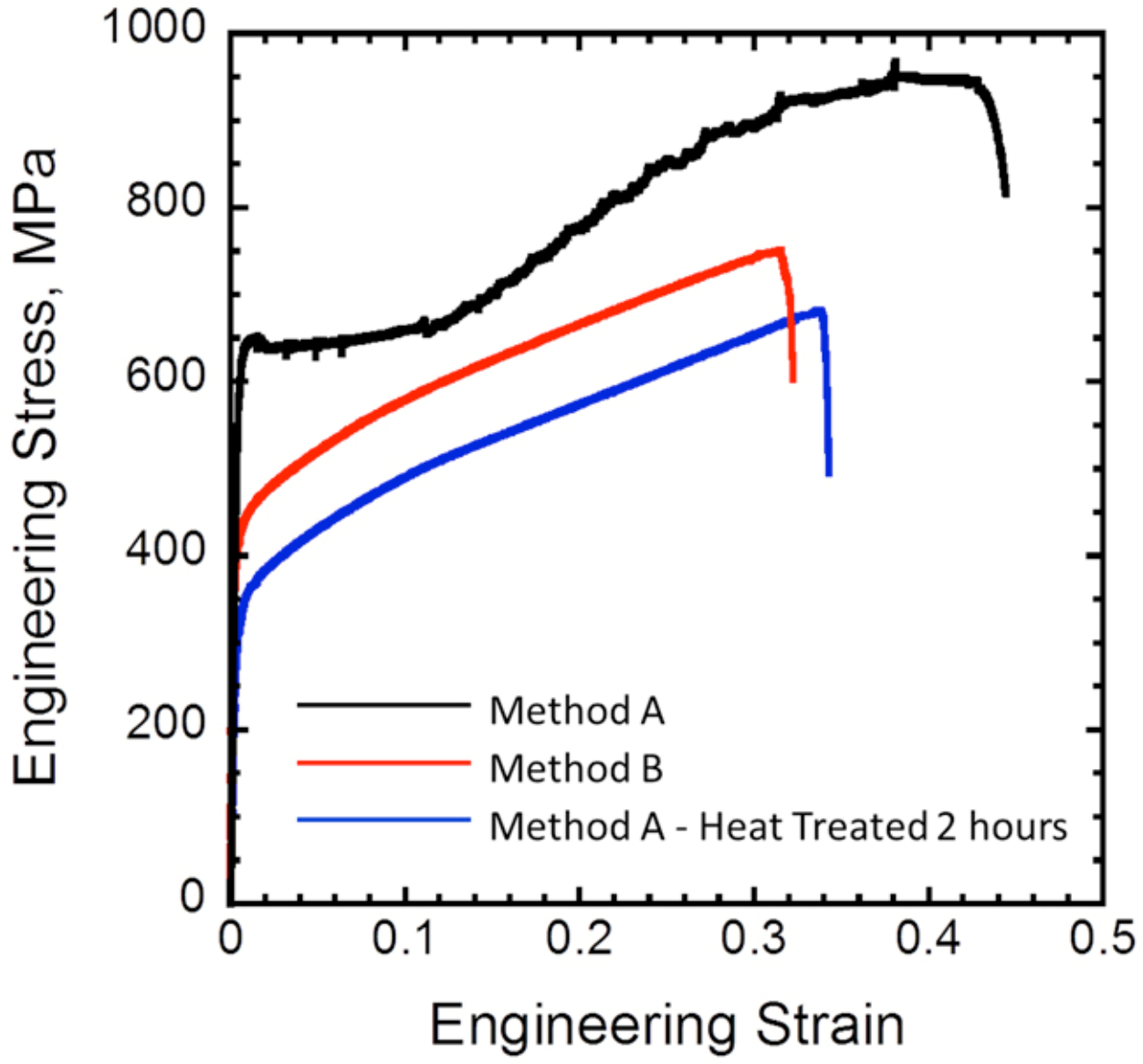

Fig. 10 – Representative tensile tests of Fe-13.92Mn-4.53Al-1.28Si-0.11C after being processed by method A (black curve); method B (red curve); and method A heat treated for 2 hours at 1123 K (850°C) (blue curve).

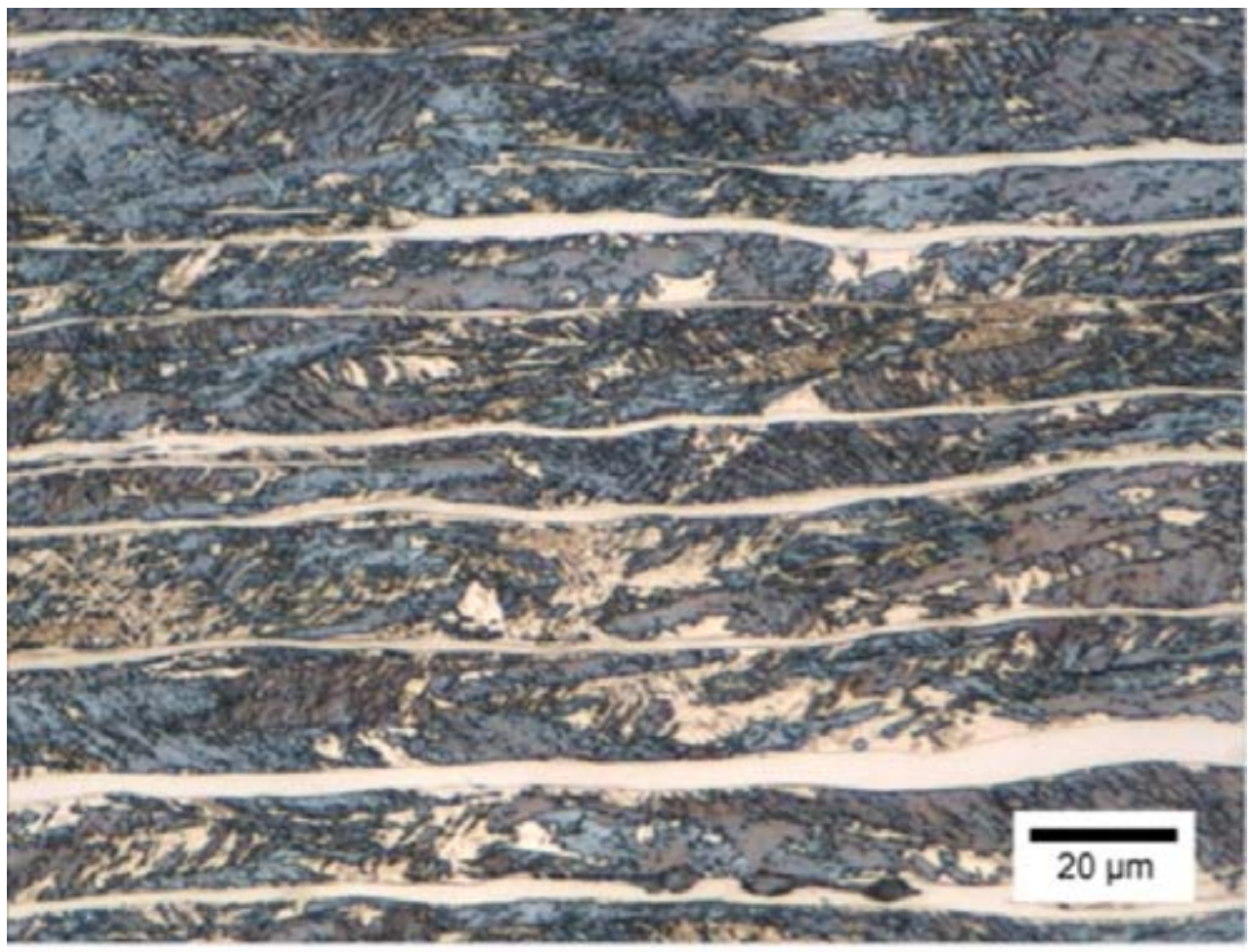

Fig. 11 – Light optical micrograph from gage section of a plate that was processed by method A and pulled in tension to failure at 970 MPa with 44% elongation.

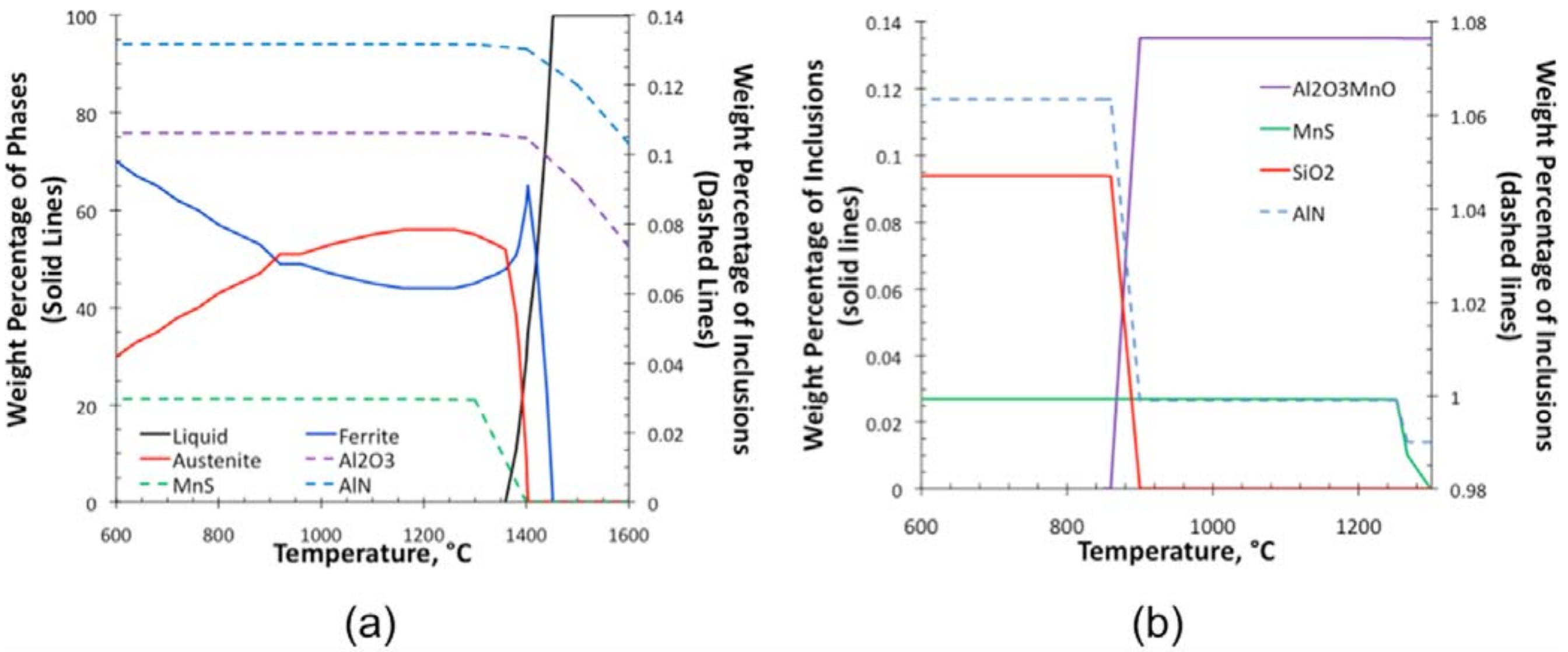

Fig. 12 – FactSage[38] predictions for nonmetallic inclusions in steels with composition (a) Fe-13.92Mn-4.53Al-1.28Si-0.045N-0.11C-0.05O-0.05S (bulk); (b) Fe-20.3Mn-0.7Al-3.2Si-1.2N-0.25C-0.05O-0.05S (last 15% to solidify).

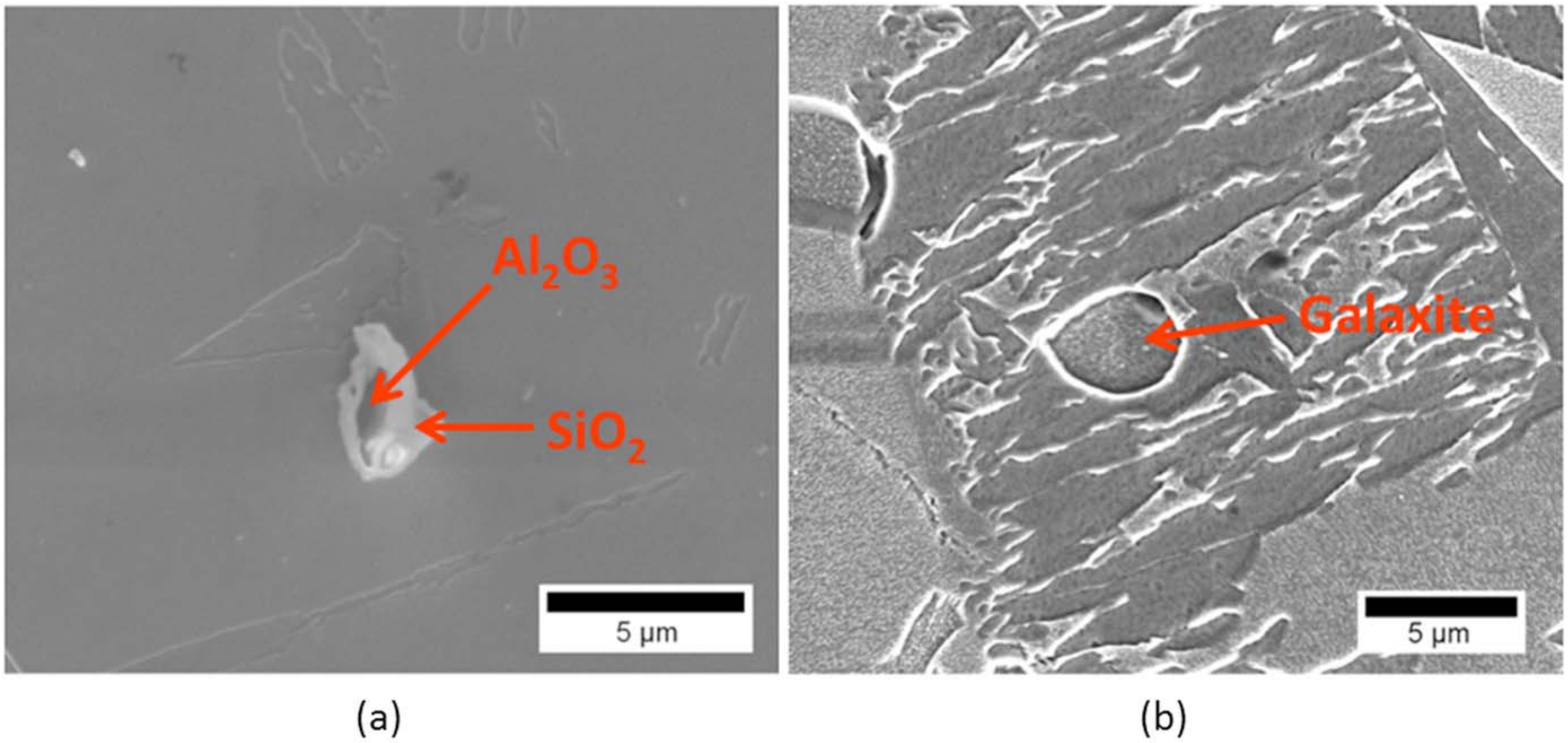

Fig. 13 – Secondary electron images of (a) silica nucleating on an aluminum oxide particle in a sample processed by method A and heat treated at 1123 K (850°C) for 47 hours and (b) galaxite observed within a cluster of acicular ferrite in a sample processed by method B and held at 1373 K (1100°C) after being heat treated at 1123 K (850°C) for 47 hours.

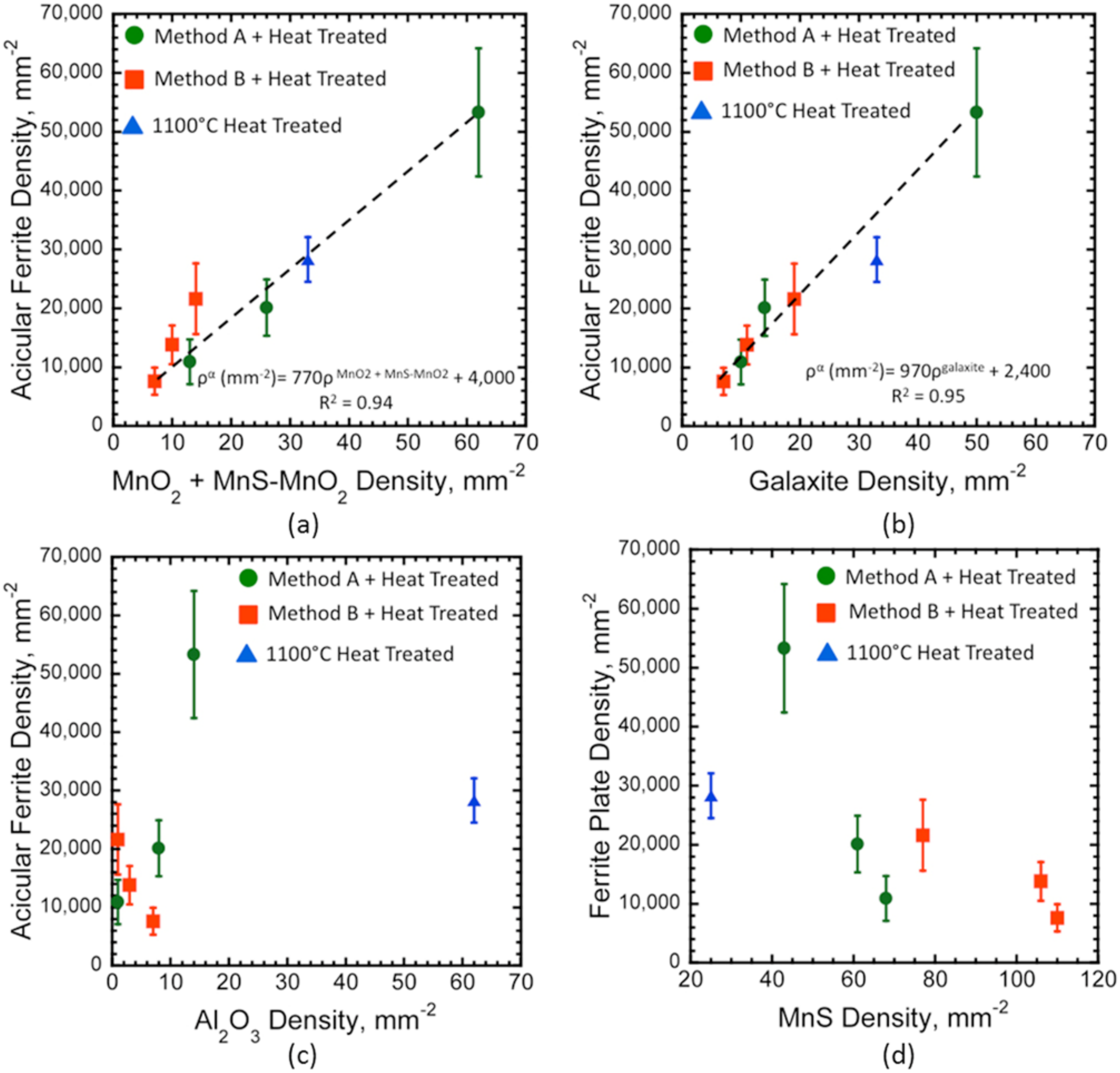

Fig. 14 – Relationships between acicular ferrite density and density of (a) MnO2 and MnO2 nucleated on MnS; (b) galaxite; (c) Al2O3; and (d) MnS for inclusions larger than 0.6 μm.

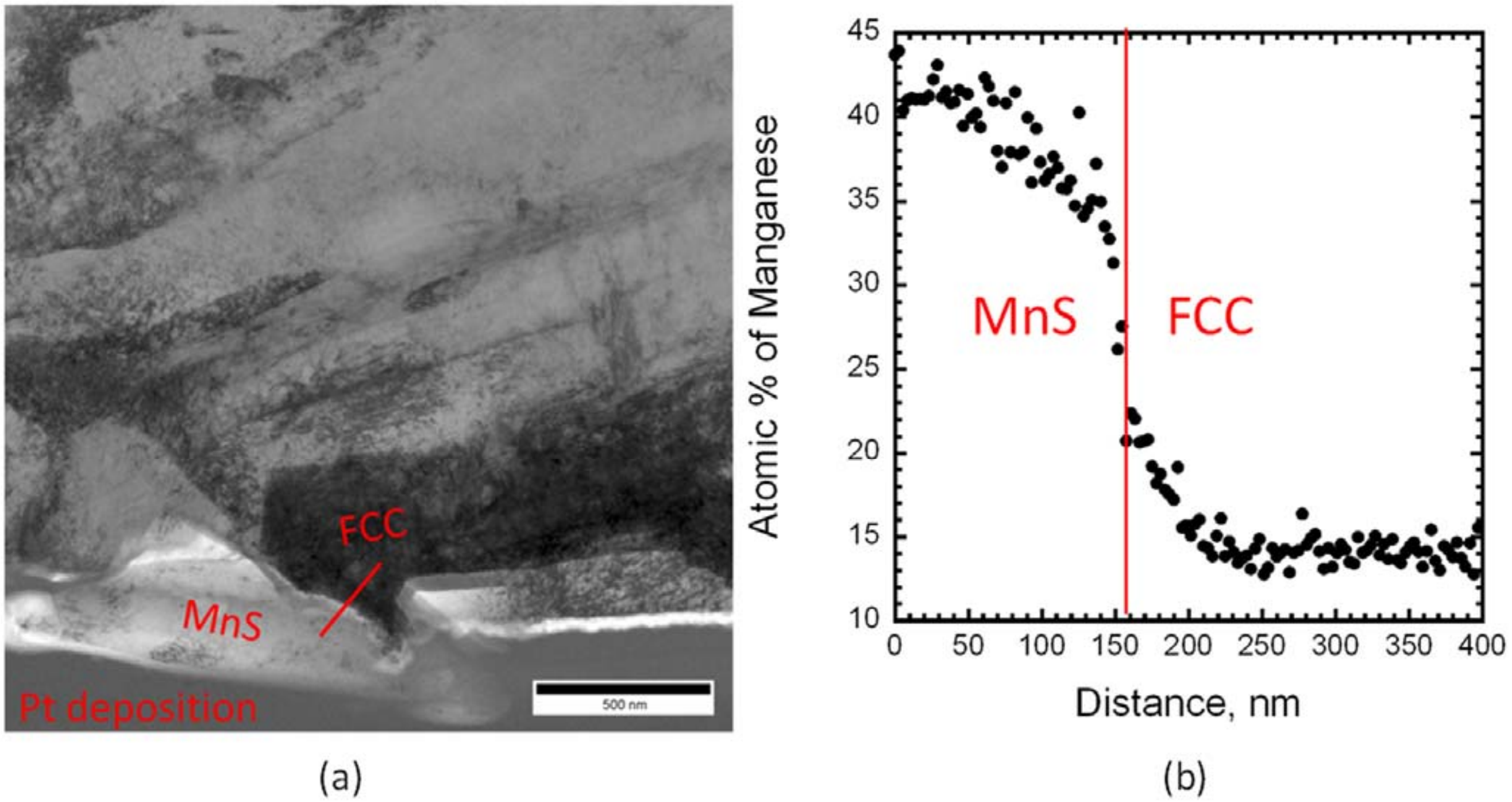

Fig. 15 – (a) TEM image showing boundary between manganese sulfide (MnS) and austenite (FCC). (b) Manganese profile across the MnS and FCC boundary. There was no depletion of manganese.

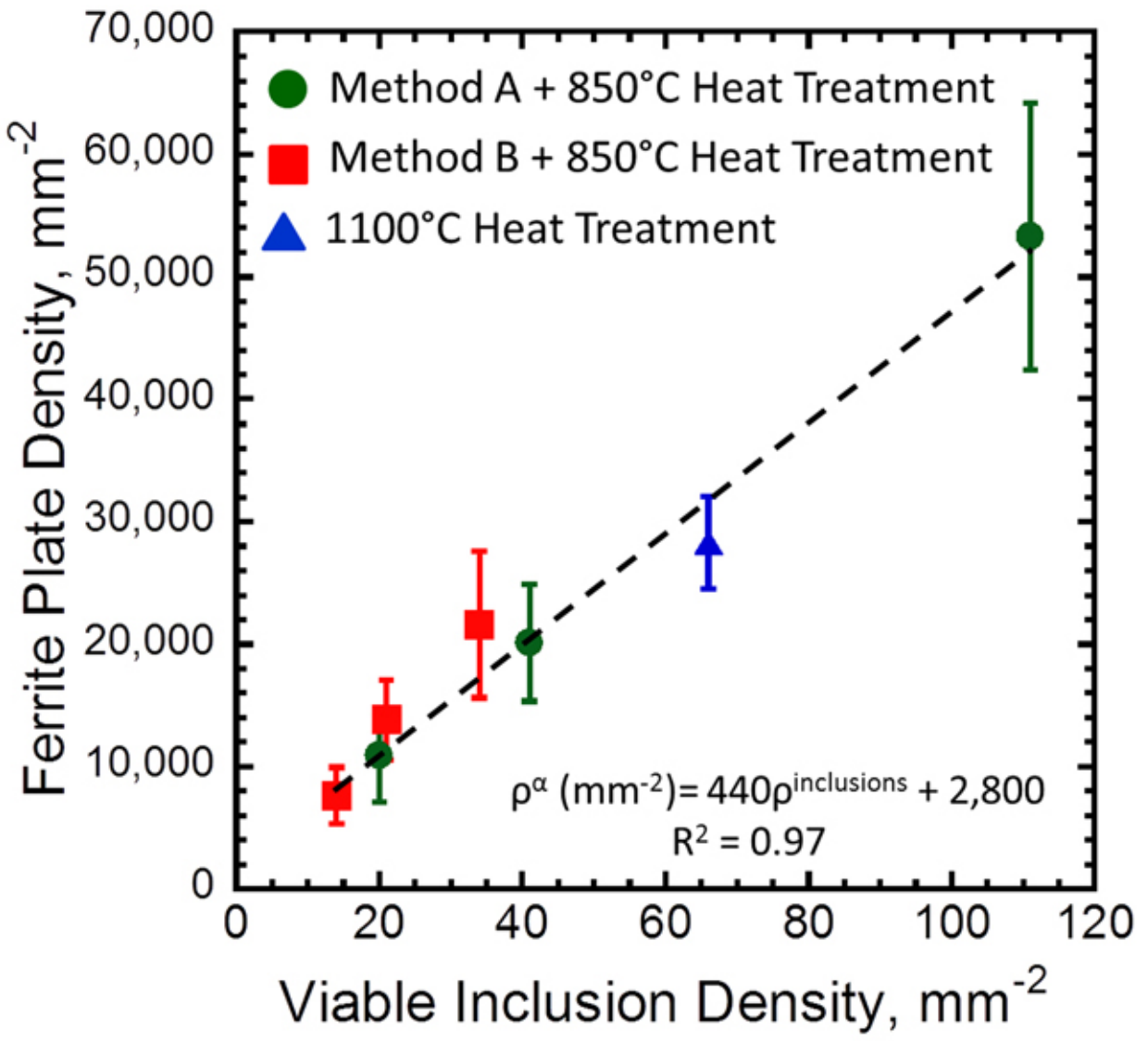

Fig. 16 – A positive relationship between acicular ferrite plate density and inclusions with low misfit that were larger than 0.6 μm was observed.

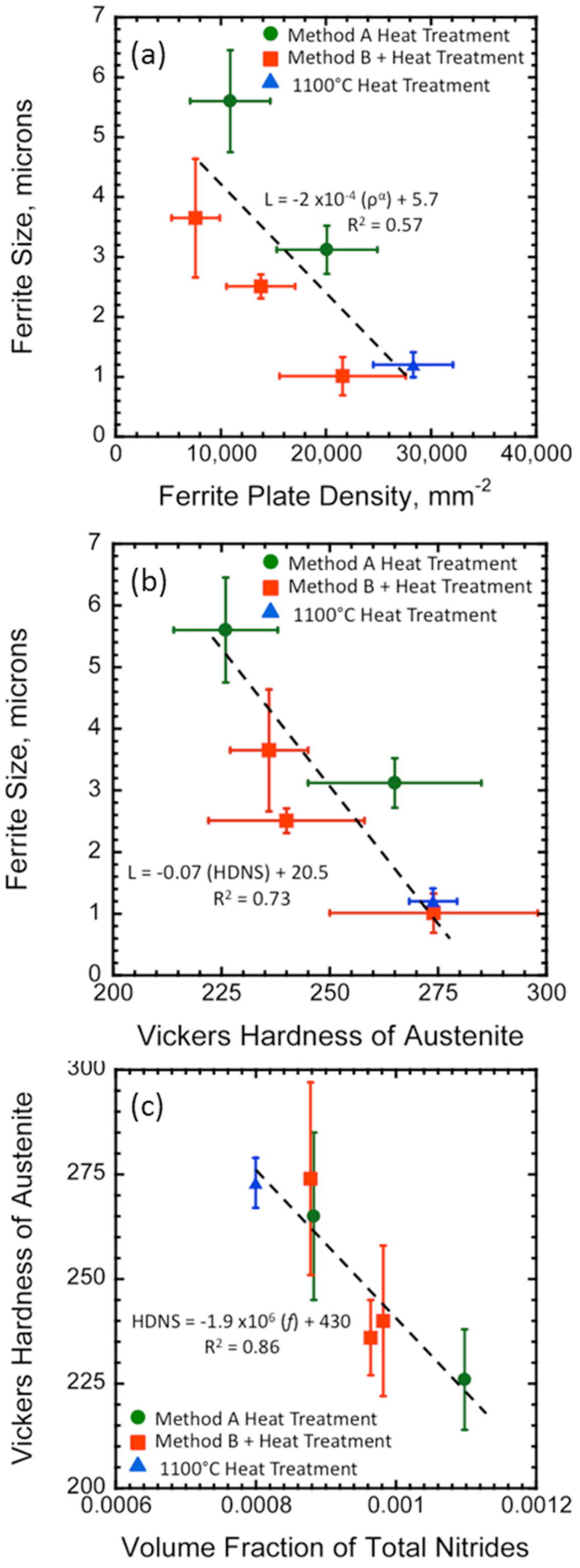

Fig. 17 – α-ferrite plate size was a function of (a) ferrite plate density and (b) austenite hardness. (c) The hardness of the austenite was a function of the total nitrides. The datum of processing by method A was removed due to this steel being in a work hardened state prior to transformation.

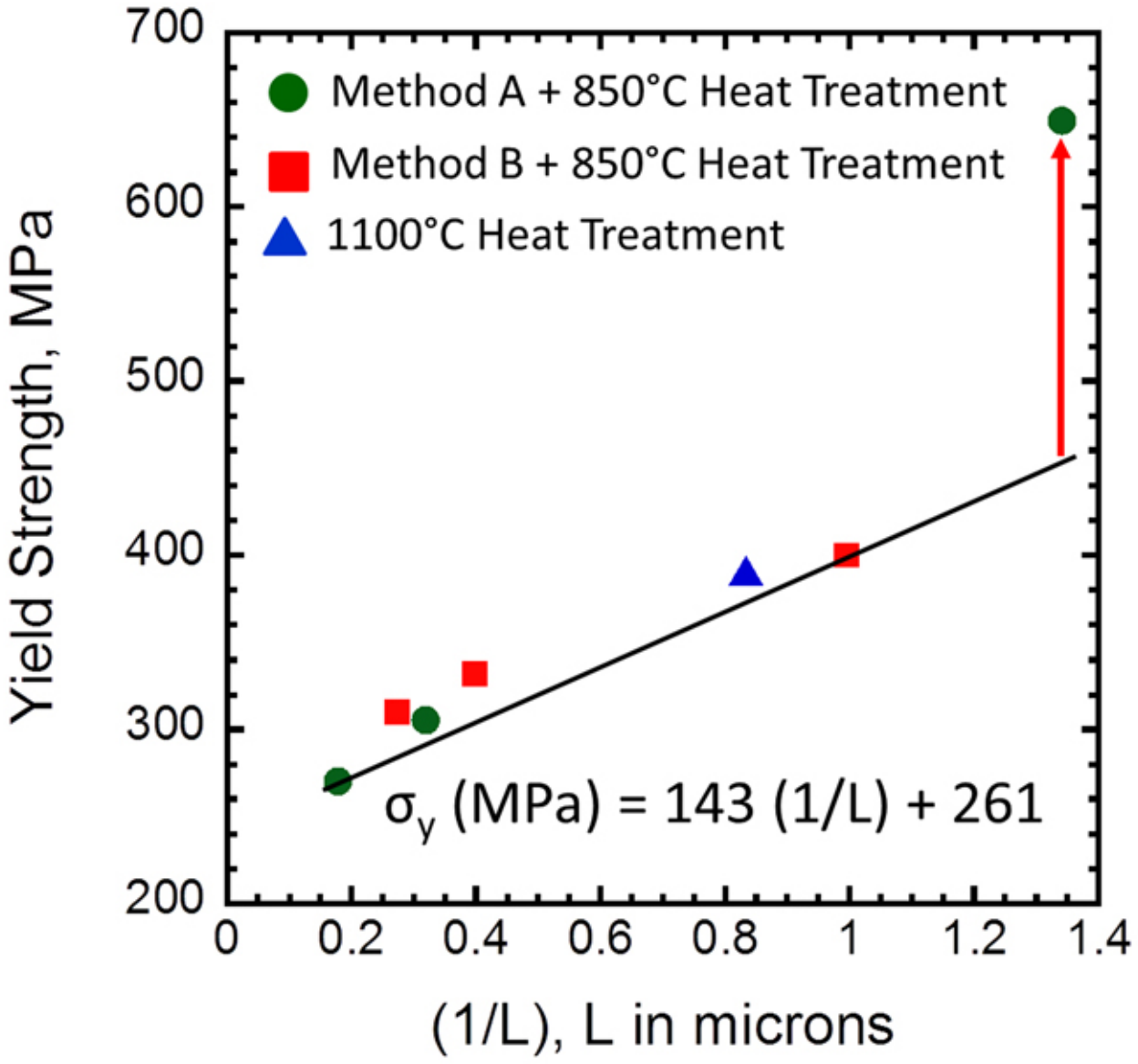

Fig. 18 – Yield strength was inversely related to the α-ferrite plate size. The plate processed by method A had retention of the dislocation structure from cold working which increased the strength of the material. This data point was excluded from determining the linear relationship between yield strength and the inverse of the size of the acicular ferrite.

LIST OF TABLE CAPTIONS

LIST OF TABLES

Table I.  Summary of Density and Size of α-Ferrite Plates and Volume Fractions of Phases

| | α-Ferrite Plate Density, $mm^{-2}$ | α-Ferrite Plate size, μm | $V_f$ of Austenite | $V_f$ of δ-Ferrite |
|---|---|---|---|---|
| Method A | 53,300 ± 10,900 | 0.75 ± 0.19 | 0.35 ± 0.06 | 0.16 ± 0.05 |
| Heat Treated at 1123 K (850 ºC) for: | | | | |
| 2 hrs | 20,100 ± 4,800 | 3.12 ± 0.40 | 0.35 ± 0.08 | 0.16 ± 0.02 |
| 47 hrs | 10,900 ± 3,800 | 5.6 ± 0.85 | 0.38 ± 0.07 | 0.22 ± 0.04 |
| Method B | 21,600 ± 6,000 | 1.01 ± 0.32 | 0.40 ± 0.07 | 0.17 ± 0.05 |
| Heat Treated at 1123 K (850 ºC) for: | | | | |
| 2 hrs | 13,800 ± 3,300 | 2.51 ± 0.20 | 0.47 ± 0.03 | 0.19 ± 0.02 |
| 47 hrs | 7,600 ± 2,300 | 3.65 ± 0.99 | 0.49 ± 0.06 | 0.25 ± 0.04 |

Table II. Summary of Average Inclusion Diameter

| | Inclusion Diameter, μm |
|---|---|
| Method A | 0.9 ± 0.2 |
| Heat Treated at 1123 K (850 ºC) for: | |
| 2 hrs | 2.2 ± 0.2 |
| 47 hrs | 2.9 ± 0.2 |
| Method B | 2.2 ± 0.2 |
| Heat Treated at 1123 K (850 ºC) for: | |
| 2 hrs | 2.2 ± 0.2 |
| 47 hrs | 2.1 ± 0.2 |

Table III. Summary of α-Ferrite Plate Density and Size and Volume Fraction of Phases

| | α-Ferrite Plate Density, $mm^{-2}$ | α- Plate size, μm | $V_f$ of Austenite | $V_f$ of δ-Ferrite |
|---|---|---|---|---|
| Method B | 21,600 ± 6,0000 | 1.01 ± 0.32 | 0.40 ± 0.07 | 0.17 ± 0.05 |
| Heated at 1123 K (850 ºC) for 47 hrs | 7,600 ± 2,300 | 3.65 ± 0.99 | 0.49 ± 0.06 | 0.25 ± 0.04 |
| Heat Treated at 1373 K (1100 ºC) | 28,300 ± 3,800 | 1.20 ± 0.21 | 0.39 ± 0.05 | 0.20 ± 0.02 |

Table IV. Summary of Microhardness

| Processing Sequence | δ, Vickers Hardness | α + γ, Vickers Hardness | γ, Vickers Hardness | α-martensite (post tensile testing), Vickers Hardness |
|---|---|---|---|---|
| Method A | 395 ± 37 | 453 ± 32 | 431 ± 23 | 530 ± 19 |
| Method A Heated at 1123 K (850 ºC) for 2 hours | 326 ± 12 | 291 ± 31 | 265 ± 20 | 526 ± 68 |
| Method B | 322 ± 18 | 332 ± 15 | 295 ± 28 | 532 ± 32 |

Table V. Summary of Mechanical Properties

| Processing Sequence | Yield Strength, MPa | Ultimate Tensile Strength, MPa | % Elongation | n |
|---|---|---|---|---|
| Method A | 637 | 970 | 44 | 0.48 |
| Method A Heated at 1123 K (850 ºC) for 2 hours | 305 | 682 | 34 | 0.37 |
| Method B | 400 | 748 | 32 | 0.35 |

Table VI. Comparison of Spacing between δ-ferrite Stringers

| | δ-ferrite Stringer Spacing, μm |
|---|---|
| Method A | 17.1 ± 1.0 |
| Heat Treated at 1123 K (850 ºC) for: | |
| 2 hrs | 16.0 ± 1.6 |
| 47 hrs | 17.3 ± 1.4 |
| Method B | 16.4 ± 2.2 |
| Heat Treated at 1123 K (850 ºC) for: | |
| 2 hrs | 16.0 ± 1.5 |
| 47 hrs | 16.3 ± 2.6 |
| Heat Treated at 1373 K (1100 ºC) for: | |
| 15 minutes | 16.3 ± 1.7 |

Table VII. Lattice Misfit between Different Inclusions and Ferrite[23, 26]

| Inclusion | Misfit % |
|---|---|
| α-MnS | 8.8 |
| $\gamma$-$Al_2O_3$ | 3.2 |
| $Al_2O_3MnO$ (Galaxite) | 1.8 |
| $MnO_2$ | 1.5 |